\documentclass[12pt,preprint]{aastex}

\begin{document}
\def\fas{\hbox{$.\!\!''$}}
\def\msun{$M_{\odot}$}
\def\Rsun{$R_{\odot}$}
\def\msunyr{$M_{\odot}~yr^{-1}$}
\def\flux{erg~cm$^{-2}$~s$^{-1}$}
\def\phflux{ph~cm$^{-2}$~s$^{-1}$}
\def\phflux2{ph~keV$^{-1}$~cm$^{-2}$~s$^{-1}$}
\def\fluxlam{erg s$^{-1}$ cm$^{-2}$\AA$^{-1}$}
\def\lum{erg~s$^{-1}$}
\def\deg{$^{\rm o}$}

\newcommand{\be}{\begin{equation}}
\newcommand{\bel}[1]{\begin{equation}\label{eq:#1}}
\newcommand{\ee}{\end{equation}}
\newcommand{\bd}{\begin{displaymath}} 
\newcommand{\bea}{\begin{eqnarray}}
\newcommand{\beal}[1]{\begin{eqnarray}\label{eq:#1}}
\newcommand{\eea}{\end{eqnarray}}
\newcommand{\eqlab}[1]{\label{eq:#1}}
\newcommand{\eqref}[1]{\ref{eq:#1}}

\newcommand{\Rs}{{R_{\rm S}}}     
\newcommand{\obs}{{\rm obs}}
\newcommand{\Teff}{T_{\rm eff}}
\newcommand{\Tinf}{T_{\infty}}
\newcommand{\thetac}{\theta_{\rm c}}
\newcommand{\muc}{\mu_{\rm c}}
\newcommand{\Linf}{L_{\infty,\rm th}}

\shorttitle{Evidence for the Event Horizon}
\shortauthors{McClintock et al.}

\title{On the Lack of Thermal Emission from the Quiescent Black Hole 
XTE~J1118+480: Evidence for the Event Horizon}

\author{Jeffrey E. McClintock, Ramesh Narayan, and George B. Rybicki}
\affil{Harvard-Smithsonian Center for Astrophysics,
Cambridge, MA 02138, USA} 
\email{jem@cfa.harvard.edu,rnarayan@cfa.harvard.edu,grybicki@cfa.harvard.edu}

\begin{abstract}

A soft component of thermal emission is very commonly observed from
the surfaces of quiescent, accreting neutron stars.  We searched with
{\it Chandra} for such a surface component of emission from the
dynamical black-hole candidate XTE~J1118+480 (=\ J1118), which has a
primary mass $M_{1} \approx$ 8 \msun.  None was found, as one would
expect if the compact X-ray source is a bona fide black hole that
possesses an event horizon.  The spectrum of J1118 is well-fitted by a
simple power-law model that implies an unabsorbed luminosity of
$L_{\rm x} \approx 3.5 \times 10^{30}$ \lum~(0.3--7 keV). In our
search for a thermal component, we fitted our {\it Chandra} data to a
power-law model (with slope and $N_{\rm H}$ fixed) plus a series of
nine hydrogen-atmosphere models with radii ranging from 9/8 to 2.8
Schwarzschild radii.  For the more compact models, we included the
important effect of self-irradiation of the atmosphere.  Because of
the remarkably low column density to J1118, $N_{\rm H} \approx 1.2
\times 10^{20}$~cm$^{-2}$, we obtained very strong limits on a
hypothetical thermal source: $k\Tinf <$ 0.011 keV and $L_{\infty,\rm
th} < 9.4 \times 10^{30}$ \lum~ (99\% confidence level).  In analogy
with neutron stars, there are two possible sources of thermal
radiation from a hypothetical surface of J1118: deep crustal heating
and accretion.  The former mechanism predicts a thermal luminosity
that exceeds the above luminosity limit by a factor of $\gtrsim 25$,
which implies that either one must resort to contrived models or, as
we favor, J1118 is a true black hole with an event horizon.  In
addition to neutron stars, we also consider emission from several
exotic models of compact stars that have been proposed as alternatives
to black holes.  As we have shown previously, accreting black holes in
quiescent X-ray binaries are very much fainter than neutron stars.
One potential explanation for this difference is the larger and hence
cooler surface of an 8 \msun~compact object that might be masked by
the ISM.  However, our upper limit on the {\it total} luminosity of
J1118 of $1.3 \times 10^{31}$ \lum~is far below the luminosities
observed for neutron stars.  This result strengthens our long-held
position that black holes are faint relative to neutron stars because
they possess an event horizon.

\end{abstract}


\section{Introduction}

In principle it is possible to detect the radiation emitted from the
surface of any ordinary astronomical body such as a planet or a star
of any kind.  On the other hand, it is quite impossible to detect any
radiation from an event horizon, which is the immaterial surface of
infinite redshift that defines a black hole.  This is unfortunate
because demonstrating the reality of the event horizon is a problem
central to physics and astrophysics.  Nevertheless, despite the
complete absence of any emitted radiation, it is possible to marshal
strong circumstantial evidence for the reality of the event horizon.
One fruitful approach is based on comparing low mass X-ray binaries
(LMXBs) that contain black hole primaries with very similar LMXBs that
contain neutron star primaries.  In the quiescent state of these
systems (McClintock \& Remillard 2004), the lack of a stellar surface
leads to predictable consequences, such as the faintness of black
holes relative to neutron stars (Narayan, Garcia, \& McClintock 1997b;
Garcia et al. 2001; Narayan, Garcia \& McClintock 2002), and also the
lack of a thermal component of emission from black holes, which is
commonly present in neutron stars (this work).  Similarly, in the
outburst state, the presence of a surface in the neutron star (NS)
systems gives rise to some distinctive phenomena that are absent in
the black hole (BH) systems: (1) type I thermonuclear bursts (Narayan
\& Heyl 2002); (2) high-frequency ($\sim1$~kHz) timing noise (Sunyaev
\& Revnivtsev 2000); and (3) a distinctive spectral component from a
boundary layer at the stellar surface (Done \& Gierlinski 2003).

In quiescence, almost all accreting neutron stars (e.g., Cen X--4, Aql
X--1, and KS 1731--260) display a soft (kT $\sim$ 0.1 keV) thermal
component of emission (see \S4.2).  The source of the thermal energy
is uncertain; for example, it may be energy liberated by the impact of
accreting matter (Narayan et al. 1997b, 2002) or crustal energy from the
star's interior (Brown, Bildsten, \& Rutledge 1998; Rutledge et
al. 2002b).  In any case, however, the observed X-ray luminosities,
temperatures and distances of these NSs indicate that one is observing
thermal emission from a source of radius $\sim10$~km, which is plainly
the stellar surface.

On the other hand, no quiescent thermal component of emission has been
reported for any of the 15 BH LMXBs (McClintock \& Remillard 2004),
which is the expected result if they possess an event horizon.  These
BH spectra are well-represented by a simple power law with photon
index $1.5 < \Gamma < 2.1$ (McClintock \& Remillard 2004).  However, a
soft component of emission might have escaped detection for several
reasons.  For example, the quiescent BH LMXBs are fainter, making a
soft component more difficult to observe.  Also, compared to the
surface of a NS, a hypothetical material surface surrounding a
$\sim10$~\msun~compact object would be larger and therefore have a
correspondingly lower surface temperature.

Herein we search for a thermal component of emission in the quiescent
spectrum of XTE J1118+480 (hereafter J1118), a BH LMXB with an
extraordinarily high Galactic latitude ($b = 62$\deg) and
correspondingly low interstellar absorption: $N_{\rm H}~\approx
1.2~\times~10^{20}$~cm$^{-2}$ (see \S2).  For this nominal column
depth, the transmission of the interstellar medium (ISM) is 70\% for
the softest X-rays (0.3 keV) that we consider.  Thus, J1118 provides a
unique opportunity to search for a soft thermal component of X-ray
emission.  For the purposes of this study, we adopt a mass for the BH
primary of $M_1 = 8$ \msun~(McClintock et al. 2001a; Wagner et
al. 2001; Orosz et al. 2004).  The quiescent X-ray luminosity is
$\approx3.5 \times 10^{30}$ \lum~ (0.3--7 keV; D = 1.8 kpc), which is
$10^{-8.5}$ of the Eddington luminosity (McClintock et al. 2003).
Both the luminosity and the photon spectral index, $\Gamma~=
2.02~\pm~0.16$, are typical for a quiescent BH LMXB with a short
orbital period, $P_{\rm orb} = 4.1$~hr (McClintock et al. 2003).  All
of the BH and NS LMXBs that we consider herein are X-ray novae
(a.k.a. soft X-ray transients) that undergo bright outbursts lasting
several months, which are followed by years or decades of quiescence.
During its outburst maximum, J1118 was exceptionally underluminous in
the 2--12 keV band, $L_{\rm x} \approx 3 \times 10^{35}$ \lum, compared
to the other BH LMXBs in outburst, $L_{\rm x} \sim 10^{38}$ \lum
(McClintock \& Remillard 2004).

In this work, we determine a strong upper limit on any soft thermal
component in the {\it Chandra} X-ray spectrum of J1118.  Using our
upper limit on this emission component, and assuming that J1118
possesses a hypothetical material surface, we set stringent
temperature and luminosity upper limits on thermal emission from this
surface for a wide range of assumed surface radii; we compare these
limits to the observed temperatures and luminosities of quiescent NS
LMXBs.  We conclude that the absence of a soft thermal component of
emission in the spectrum of J1118 rules strongly against the presence
of a material surface and hence argues for the existence of an event
horizon.

This work is organized as follows.  In \S2 we examine the central
question of the column density to J1118, and in \S3 we discuss the
observations, data analysis and model-fitting techniques.  The
development and computation of stellar atmospheric models appropriate
to a compact and massive star are presented in Appendix A; the models
include the effects of self-irradiation for stars so compact as to lie
within their own photon spheres.  In \S4, upper limits on the
temperature and luminosity of a thermal component from J1118 are
summarized and discussed, and these results are compared to the
thermal spectra observed for neutron stars.  In \S5, we interpret the
absence of thermal emission form J1118 in terms of two conventional
sources of thermal emission from NSs; in addition we consider emission
from exotic models of massive, compact stars that have been proposed
as alternatives to BHs.  Our conclusions are summarized in \S6.  J1118
and the other BHs referred to throughout this work are among the 18
dynamically-confirmed BHs; for a review of the properties of these
massive, compact X-ray sources, see McClintock \& Remillard (2004).

\section{Interstellar Absorption}

Knowledge of the interstellar absorption is crucial to the
determination of the limits on the temperatures and luminosities of
the compact objects that are modeled in this work.  In this regard,
the very high Galactic latitude of J1118 ($b = 62^{\rm o}$) and
therefore low interstellar absorption has already been noted (\S1).
The location of J1118 close to the Lockman Hole (Lockman, Jahoda \&
McCammon 1986) indicates an expected column density of $N_{\rm H}~=
0.5-1.5 \times 10^{20}$~cm$^{-2}$ (see Hynes et al. 2000).  As another
indicator of low $N_{\rm H}$, J1118 is the only X-ray nova that was
observable by the {\it Extreme Ultraviolet Explorer (EUVE)} during its
9-year mission (Hynes et al. 2000), which is especially remarkable
given the faintness of the X-ray source during outburst (\S1).  The
good quality of the EUV spectrum (0.10--0.19 keV) obtained for J1118
(Hynes et al. 2000) strongly attests to the low interstellar
absorption, especially when one considers that the transmission of the
ISM is only $\approx 10^{-3}$ at 0.10 keV for a nominal column depth
of $N_{\rm H}~= 1.2 \times 10^{20}$~cm$^{-2}$ (see below).

Plainly, the EUV fluxes corrected for interstellar absorption are
extremely sensitive to the value assumed for $N_{\rm H}$ (Hynes et
al. 2000).  This fact plus a very complete set of simultaneous
observations of J1118 in outburst obtained using UKIRT, {\it EUVE},
{\it HST}, {\it Chandra} and {\it RXTE} have allowed strong
constraints to be placed on $N_{\rm H}$.  The one reasonable
assumption that must be made is that the HST data (1.2--10.7 eV),
which are very {\it insensitive} to the choice of $N_{\rm H}$,
represent the Rayleigh-Jeans portion of a cool ($kT \approx 24$ eV)
accretion disk spectrum, whereas the EUVE data (100--190 eV)
constitute the Wien portion of this disk spectrum (McClintock et
al. 2001b).  (The X-ray data extending from 0.24--160 keV 
correspond to a separate
power-law component of emission; McClintock et al. 2001b).  The
preferred range of $N_{\rm H}$ for which the {\it HST} and {\it EUVE}
data harmonize to form an accretion disk spectrum is $N_{\rm H}~=
1.0-1.3 \times 10^{20}$~cm$^{-2}$.  This determination of the column
density strongly rules against $N_{\rm H}~<
0.75 \times 10^{20}$~cm$^{-2}$ or $N_{\rm H}~>
1.6 \times 10^{20}$~cm$^{-2}$.  These results have been corroborated
by a recent reanalysis of these data and an analysis of much
additional multiwavelength data by Chaty et al. (2003) who conclude
that $N_{\rm H}~= 0.80-1.30 \times 10^{20}$~cm$^{-2}$ with a preferred
value of $N_{\rm H}~= 1.1 \times 10^{20}$~cm$^{-2}$.

In assessing the hydrogen column density to J1118, Frontera et
al. (2001) consider results based on radio measurements of H I (Dickey
\& Lockman 1990), the detection of weak interstellar Ca II lines
(Dubus et al. 2001), and fits to their {\it Beppo-SAX} X-ray
spectra.  They conclude that the ``most likely range'' of $N_{\rm H}$
is $(1-1.5) \times 10^{20}$ cm$^{-2}$.  A subsequent {\it Beppo-SAX}
observation indicated a value of $N_{\rm H}$ near the low end of this
range (Frontera et al. 2003).  In conclusion, based on the three
independent lines of argument sketched above --- namely, the radio
results (see Hynes et al. 2000; Dickey \& Lockman 1990), the
multiwavelength spectral results (McClintock et al. 2001b; Chaty et
al. 2003), and the {\it BeppoSax} spectral fits (Frontera et al. 2001,
2003) --- {\it we adopt $N_{\rm H}~= 1.2 \times 10^{20}$~cm$^{-2}$ as
the most likely value of the column density, and $N_{\rm H}~<
1.6 \times 10^{20}$~cm$^{-2}$ as a conservative upper limit.}  As we
show below, our results are scarcely affected by uncertainties in the
column density because the transmission of the ISM is very high even
at the lowest energies considered here (e.g., 70\% at 0.3 keV for 
$N_{\rm H}~= 1.2 \times 10^{20}$~cm$^{-2}$).

\section{Observations, Data Analysis, and Model Fitting}

The X-ray data were obtained with the Advanced CCD Imaging
Spectrometer (ACIS; Garmire et al. 1992) onboard {\it The Chandra
X-ray Observatory} on 2002 January 11 and 12 UT.  The data are
identical to those described in McClintock et al. (2003; hereafter
MNG03).  MNG03 used only a 45.8 ks subset of the {\it Chandra} data
that was obtained on January 12 during simultaneous observations made
with the {\it HST} and the MMT.  In this work we include an additional
7.7 ks of data that were obtained one day earlier (see MNG03 for
details); the analysis of these data was identical to the analysis
described previously for the larger data set, and we refer the reader
to the discussion in MNG03 for a number of the details, which we omit
here.  The resulting pair of pulse height spectra and their response
files were combined using FTOOLS (Arnaud \& Dorman 2003) which yielded
a single spectrum and response file with a net exposure time of
53.5~ks and a total of 89 source counts. The predicted number of
background counts in the source extraction circle was small, 1.0
counts, and we neglected the background in our spectral analysis.
As before, the response file is corrected for the ongoing degradation
in the ACIS-S low energy quantum efficiency (MNG03).

The source counts were binned into 12 bins, each with 7--8 counts per
bin.  As in MNG03, we fitted the pulse height spectrum to several
single-component spectral models (Arnaud \& Dorman 2003) with
interstellar absorption (Balucinska-Church \& McCammon 1992), fixing
the column density to the value determined in outburst: $N_{\rm
H}~=~1.2~\times~10^{20}$~cm$^{-2}$ (see \S2).  For a power-law model
we find a photon index $\Gamma = 1.92\pm 0.16$ ($\chi^{2}_{\rm \nu} =
1.01$ for 10~dof).  This fit is based on $\chi^{2}$ statistics with a
conventional $\sqrt N$ weighting function, and the results are
entirely consistent with those presented in MNG03.  As in MNG03, we
use this power-law model to represent the observed spectrum throughout
this work.  However, we note that we also could have used a
bremsstrahlung model with $kT = 1.63$ keV ($\chi^{2}_{\rm \nu} = 0.98$
for 10~dof).  However, some other common thermal models give poor or
unacceptable fits to the data.  For example, the blackbody model and
the Raymond-Smith model (with cosmic abundances) give values of
reduced $\chi^{2}$ of 1.7 and and 2.8 (10 dof), respectively.
However, even these models can be fitted satisfactorily if one allows
$N_{\rm H}$ to vary freely.

In this work, we adopt the following conservative approach.  Because
we have only 7--8 counts per bin, we use the Gehrels' weighting
function ($1 + \sqrt{N + 0.75}$) in conjuction with $\chi^{2}$
statistics, which provides a good approximation to Poisson statistics
(Gehrels 1986; Arnaud \& Dorman 2003).  Using this approach, we refitted
the binned data with the power-law model and found the same value for
the photon index, although the error was increased somewhat (38\%), as
expected: $\Gamma (\rm{Gehrels}) = 1.92 \pm 0.23$ ($\chi^{2}_{\rm \nu}
= 0.50$ for 10~dof).  {\it The Gehrels' weighting function gives larger
errors and therefore more conservative upper limits on surface
temperature and luminosity; we therefore use this method throughout
this work.}  

We next fitted the {\it Chandra} data to a composite model that
included the power-law component just discussed plus a
hydrogen-atmosphere spectral model; the calculation of these
atmospheric models is described in Appendix A.  Our models are similar
to those used routinely to model the spectra of quiescent neutron
stars (e.g., Zavlin, Pavlov, \& Shibanov 1996; Rutledge et al. 1999).
However, we extend our models to much larger values of mass and
radius, and we also consistently include the effects of
self-irradiation.  Using the theory and methods detailed in Appendix A
and adopting $M(J1118)$ = 8~\msun~(\S1), we computed a grid of spectra
on each of nine different surfaces with radii relative to the
Schwarzschild radius $\Rs$ ranging from $\log R/\Rs$ = 0.05 to 0.45 in
equal logarithmic steps of 0.05.  Our smallest radius corresponds to
the Buchdahl limit of $R_{\rm min} = 9/8\, \Rs$, which is the minimum
allowable radius for a stable star in general relativity (Buchdahl
1959; Weinberg 1972).  At each radius, we computed 25 photon spectra
($E =$ 0.01--10.0 keV) which correspond to temperatures of $\log \Tinf$
= 5.00, 5.01, 5.02, $\ldots$, 5.25, where the temperature at infinity
$\Tinf$ is related to the effective temperature at the surface of the
star by $\Tinf = \Teff/(1+z)$.  The models were cast in the form of
nine one-parameter ($\Tinf$) FITS table models, one for each radius,
using a Fortran program (onepar.f) that was kindly provided by Keith
Arnaud.  The table models so derived were used within XSPEC (Arnaud \&
Dorman 2003) to model a hydrogen atmosphere for the range of radii and
temperatures described above.

At each radius, we determined an upper limit on $\Tinf$ of a thermal
component as follows.  We first fitted the 12-channel pulse-height
spectrum using Gehrels' weighting to a model that includes
interstellar absorption and a simple power-law component.  In this
case, however, we fixed the column depth at its estimated maximum
value of $N_{\rm H}~=~1.6~\times~10^{20}$~cm$^{-2}$ (\S2) in order to
obtain the most conservative (i.e., largest) upper limits on
$\Tinf$~and $L_{\infty,\rm th}$ (the thermal component of luminosity
as measured at infinity).  In this way we obtained
$\Gamma$(Gehrels)~$= 1.97~\pm~0.24$ with a flux at 1 keV = ($1.70 \pm
0.55) \times 10^{-6}$~\phflux2 ($\chi^{2}_{\rm \nu}~=~0.47$ for
10~dof).

We next created a composite source model by fixing the power-law
component using the just-mentioned values of $\Gamma$ and $N_{\rm H}$
and by adding in turn each of the nine atmospheric models.  Since the
normalization of the thermal component is determined by the source
radius $R$ and distance $D = 1.8$ kpc, there remains only a single
free parameter, $\Tinf$.  Accordingly, we determined a 99\% confidence
level upper limit on $\Tinf$~by successively increasing its value
until the total $\chi^2$ increased by 6.63 (Lampton, Margon, \& Bowyer
1976).

\section{Results}

\subsection{Limits on Temperature and Luminosity}

We determined upper limits at the 99\% confidence level on $\Tinf$ by
the method described in \S3; these limits are given for each of the
nine H-atmosphere models in column 7 of Table 1.
Columns 2 through 6 give the radius, surface redshift, critical cosine
$\muc$ (defined in Appendix A), $\log g$ (surface gravity), and $\log
\Teff$ for each model.  The limits on $\Tinf$ are shown plotted vs.\
radius in Figure \ref{fig:1}\placefigure{fig:1} (filled circles).  The
highest temperature limits occur for Models \ 3 and 4 with $R \approx
1.4-1.6 ~\Rs$: $\Tinf \leq 126,200$~K.  We emphasize that these are
strong limits on $\Tinf$ for two reasons: First, in fitting the data
we have adopted the Gehrels' weighting function which gives larger
errors and hence more conservative limits than the conventional
weighting function (\S3).  And second, these limits are based on the
estimated maximum allowable value of the interstellar column density,
$N_{\rm H}~= 1.6 \times 10^{20}$~cm$^{-2}$ (\S2).  Furthermore, as
illustrated in Figure \ref{fig:1}, the temperature limits depend only
weakly on $N_{\rm H}$ because the ISM is optically thin even at the
lowest energies (\S1).

Note that the effective temperature as measured at the surface,
$\Teff$ (Table 1, column 6), varies inversely with radius, reaching a
maximum at the Buchdahl limit: $R = 9/8\, \Rs$.  Columns 8 and 9 give
upper limits on the luminosities at infinity $L_{\infty,\rm th}$ and
at the surface of the star $L_{\rm R}$, respectively, where $L_{\rm R}
= 4\pi R^2 \sigma T_{\rm eff}^{4}$ and $L_{\infty,\rm th} = L_{\rm
R}/(1+z)^2$; these limits are plotted in Figure
\ref{fig:2}.\placefigure{fig:2} The limiting value of $L_{\infty,\rm
th}$, which is of most interest, varies by only $\approx 40$\% over
the range of radii considered; it reaches a maximum value for the most
compact configuration, namely Model~1: $L_{\infty,\rm th} < 9.4 \times
10^{30}$~\lum~(99\% confidence level).

Because the compact objects modeled here are relatively large and
cool, it might be thought that useful limits could also be achieved
through observations in the UV.  However, as shown in Figure
\ref{fig:3} \placefigure{fig:3}, this is not possible.  The figure
shows the relationship of six atmospheric models to a multiwavelength
spectrum of J1118, which was published previously (MNG03).  The models
shown completely bracket the ranges of temperatures and radii
considered in the text and in Table~1. The heavy horizontal line on
the right is the best-fit model X-ray spectrum and indicates that {\it
Chandra} is capable of detecting the surface emission for
log$T_{\infty} = 5.15$, but not for the cooler models with
log$T_{\infty} = 5.05$, as expected given the temperature limits
summarized in Table~1.  On the left of the plot is shown the
UV/optical spectrum.  Of most interest here is the HST FUV spectrum,
which is centered at $\nu \sim 2 \times 10^{15}$ Hz, and plunges
downward to log${\nu}F_{\nu} \sim -14.1$.  These data were obtained in
a 14 ks observation with STIS using the G140L grating.  A hard limit
on the sensitivity of this observation is indicated by the upward
arrow.  The models in question are all much fainter than the UV limit
indicated.  We conclude that HST cannot provide useful constraints on
the models. We also conclude that the observed UV/optical emission
cannot be due entirely or in part to thermal emission from the surface
of the compact object.  For a discussion of the origin of the
UV/optical emission, see MNG03.

\begin{deluxetable}{ccccccccc}\label{tab:1}

\tablewidth{0pt}  

\tablecaption{Upper Limits\tablenotemark{a} on Temperature and
Luminosity\tablenotemark{b}}

\tablehead{
  \colhead{Model} & \colhead{$\log R/R_{\rm S}$} & \colhead{$1+z$} &
  \colhead{$\muc$} & \colhead{$\log g$} & \colhead{$\log\Teff$} &
  \colhead{$\log\Tinf$} & \colhead{$\log L_{\infty,\rm th}$} &
  \colhead{$\log L_{\rm R}$}  }

\startdata

1  &  0.0512 &  3.000  &  0.6383 &  14.654&  5.557 & 5.080  & 30.974 & 
31.928 \\
				              	                
2  &  0.10 &  2.205  &  0.3522 &  14.423&  5.436 & 5.092  & 30.855 & 
31.541 \\
 				              	                
3  &  0.15 &  1.851  &  0.1094 &  14.246&  5.367 & 5.100  & 30.833 & 
31.368 \\
				              	                
4  &  0.20 &  1.646  &  0.0    &  14.096&  5.317 & 5.101  & 30.836 & 
31.269 \\
				              	                
5  &  0.25 &  1.512  &  0.0    &  13.959&  5.277 & 5.098  & 30.848 & 
31.207 \\
				              	                
6  &  0.30 &  1.416  &  0.0    &  13.830&  5.243 & 5.092  & 30.868 & 
31.170 \\
				              	                
7  &  0.35 &  1.344  &  0.0    &  13.708&  5.213 & 5.084  & 30.893 & 
31.150 \\
				              	                
8  &  0.40 &  1.289  &  0.0    &  13.589&  5.186 & 5.075  & 30.922 & 
31.142 \\
				              	                
9  &  0.45 &  1.245  &  0.0    &  13.474&  5.161 & 5.066  & 30.953 & 
31.144

\enddata

\vspace*{-0.2in}

\tablecomments{~All quantities are in cgs units.}

\tablenotetext{a}{~Limits are at the 99\% level of confidence and
based conservatively on a column density of $N_{\rm H}~=
1.6~\times~10^{20}$~cm$^{-2}$ (\S2) and Gehrels' weighting (\S3).}

\tablenotetext{b}{~Legend: $R$, stellar radius, $\Rs$, Schwarzschild
radius, $z$, surface redshift, $\muc$, cosine of critical angle, $g$,
surface gravity, $\Teff$, effective temperature at stellar surface,
$\Tinf$, temperature at infinity, $\Linf$, thermal component of
luminosity~at infinity, $L_{\rm R}$, luminosity at stellar surface.}

\end{deluxetable}

\subsection{Comparison with the Predominantly Thermal Spectra of
Neutron Stars}

The spectrum of J1118, with its absence of a thermal component,
contrasts sharply with the 0.5--10 keV spectra of quiescent NS LMXBs.
A soft thermal component is dominant (i.e., comprises $\gtrsim 50$\%
of the total flux) in the 0.5--10 keV band in nearly all NS spectra,
although a fainter power-law component is also often present.  When
the thermal component is fitted with a NS H-atmosphere (NSA) model,
the derived temperatures are in the range $kT_{\infty} = 0.05-0.3$ keV
($\log T_{\infty} =$ 5.8--6.5) and the source radii are consistent
with surface emission from a NS with a $\sim 10$ km radius.  The
following ten accreting NSs are known to have such a dominant, soft
emission component: 47 Tuc X5 and X7 (Heinke et al. 2003a); Cen X--4
(Campana et al. 2000; Rutledge et al. 2001a); 4U 1608--52 (Asai et
al. 1996); MXB 1659--29 (Wijnands et al. 2003bc); RX~J170930.2--263927
(Jonker et al. 2003a); KS 1731--260 (Wijnands et al. 2001; Wijnands et
al. 2002); X1745--203 (NGC 6440 CX1; in't Zand 2001); Aql X--1
(Rutledge et al. 2001b); and 4U~2129+47 (Nowak, Heinz, \& Begelman
2002).

For two other NSs, EXO 1745--248 and SAX J1810.8-2609, the thermal
component is less apparent, but it may well be present.  For the
latter source, Jonker, Wijnands, \& van der Klis (2003b) found that
their {\it Chandra} data are well-fitted by a simple power-law model
with $N_{\rm H}~= 3.3~\times~10^{21}$~cm$^{-2}$ and that the NSA and
other one-component thermal models provided much poorer fits.
However, when they fitted their data with an NSA plus power-law model,
the reduced $\chi^{2}$ decreased further (from 0.80 to 0.65 for 10
dof), and the inferred properties of the thermal component were
typical of the canonical systems listed above: $kT_{\infty} \approx
0.07$ keV with the thermal component contributing about half of the
unabsorbed flux.  For EXO 1745--248 in Terzan 5, the {\it Chandra}
ACIS-S data can be fitted by a simple power-law model; furthermore,
the fitted value of the very large column density, $N_{\rm H}~=
1.3~\times~10^{22}$~cm$^{-2}$, agrees with an independent estimate
(Wijnands et al. 2003a).  However, the authors conclude that a thermal
component of emission may contribute as much as 10\% of the total
emission (0.5--10 keV) with an NSA temperature as high as $kT_{\infty}
= 0.10$ keV. 

An instructive comparison object to consider is the millisecond X-ray
pulsar SAX J1808.4-3658, which lacks a detectable soft component of
emission and has a moderate column density, $N_{\rm H}~=
1.3~\times~10^{21}$~cm$^{-2}$.  Its luminosity is only a factor of
several times greater than the luminosity of J1118; furthermore, the
limit on its bolometric thermal luminosity (Campana et al. 2002) is a
factor of $\approx 5$ less than the limits we have set on the thermal
luminosity of J1118 (Fig. 2a).  However, the {\it temperature limit}
that can be placed on any hypothetical thermal component in J1118 is
far more stringent because of its minimal interstellar column: $N_{\rm
H}~= 1.2~\times~10^{20}$~cm$^{-2}$ (see below).

In the case of SAX J1808.4-3658, Campana et al. (2002) conclude that a
power-law fit alone can adequately describe their data.  They set
an upper limit on the flux of a hypothetical blackbody component that
is $<$ 7\% of the power-law flux in the 0.5--10 keV band for assumed
blackbody temperatures in the range $kT_\infty = 0.1-0.3$ keV
(unabsorbed fluxes; 90\% confidence).  Our corresponding limit on the
ratio of the unabsorbed fluxes in an 0.5--10 keV band) for J1118 is a
factor $\approx$ 12 lower, namely, 0.6\% for Model~4 (Table~1).  It is
also important to note that the $<$ 7\% limit set for SAX~J1808.4-3658
is based on a measurement of (at most) 17\% of the bolometric flux for
a $kT = 0.10$ keV blackbody (83\% of the flux is absorbed in the ISM).
Furthermore, for slightly lower and quite realistic neutron-star
surface temperatures (e.g., $k\Tinf = 0.05$ keV) the fraction of the
flux that is observable becomes extremely small (2.1\%).  Thus, a soft
but luminous thermal component may be present in SAX J1808.4-3658 that
is masked by the ISM.  On the other hand, for the order-of-magnitude
lower column density of J1118, the ISM transmits 67\% (31\%) of the
bolometric flux of a $k\Tinf = 0.10~(0.05)$ keV blackbody thereby
making a cool thermal component far easier to detect; yet none is seen
in J1118. 

Dominant soft X-ray emission is also known for 18 other X-ray sources
located in globular clusters that are thought to be quiescent LMXBs
containing NSs (Heinke et al. 2003b).  However, these objects have not
been observed in outburst nor have they exhibited type I bursts;
moreover, about half of these objects are very faint.  Consequently,
although these sources attest to the ubiquity of thermal surface
emission from NSs, they are less instructive than the confirmed and
well-studied NSs cited and discussed above.

In summary, for most NSs the thermal component is dominant; in a few
examples the thermal component is weaker but still may contribute
7--10\% of the total 0.5--10 keV flux.  In all cases where the thermal
component is detected, the NSA temperature is $k\Tinf \sim 0.1$
keV.  {\it We emphasize that the corresponding temperature limit on
thermal emission from J1118 is incomparably stronger because of the
source's very low column density: log $\Tinf < 5.10$, $k\Tinf < 0.011$
keV (Table 1)}.  This limit corresponds to a maximum thermal
contribution in the 0.5--10 keV band of $< 0.6$\% (99\% confidence
level).

\section{Discussion}

We have seen in the previous sections that XTE J1118+480 in quiescence
has no detectable component of thermal surface emission; a
conservative 99\% confidence level upper limit on the luminosity in
such a component is $L_{\infty,\rm th} < 9.4\times 10^{30} ~{\rm
erg\,s^{-1}}$.  In comparison, quiescent NS LMXBs have significantly
larger luminosities, typically $L_{\infty, \rm th} \sim {\rm few}
\times 10^{32}$ to ${\rm few}\times 10^{33} ~{\rm erg\,s^{-1}}$ (see
references in \S4.2).  Before trying to interpret this difference, we
need to decide what causes the surface emission in NS LMXBs.  Two
possibilities have been discussed in the literature: deep crustal
heating, and accretion.  Both require the accreting star to possess a
surface.  We discuss the two possibilities in the following
two subsections.

\subsection{Thermal Surface Emission from Deep Crustal Heating}

Brown et al. (1998) showed that gas accreting onto the surface of a NS
releases heat energy through nuclear reactions deep within the crust.
These deep crustal reactions occur at a pressure of
$\sim10^{30}-10^{31} ~{\rm dyne\,cm^{-2}}$.  The diffusion time from
this depth to the surface of the NS is on the order of $1-10$ years,
which is much longer than the typical duration of an accretion
outburst in a transient LMXB.  Therefore, the deep crustal heat energy
escapes mostly during the quiescence phase of the systems and is
expected to be observed as a steady thermal flux from the NS surface.

Brown et al. (1998) estimate that about 1.45 MeV of energy is released
by the deep crustal reactions per accreted nucleon.  Since the
accretion luminosity is also proportional to the number of nucleons
accreted, there is a direct proportionality between the fluence
$S_{\rm acc}$ due to accretion and the fluence $S_{\rm dch}$ due to
deep crustal heating:
\begin{equation}
S_{\rm acc} = 135 {\eta\over 0.2} S_{\rm dch}, \label{accdch}
\end{equation}
where $\eta$ is the radiative efficiency of the accretion flow:
$L_{\rm acc} = \eta\dot M c^2$.  Brown et al. (1998) show that the
quiescent luminosity predicted from deep crustal heating agrees rather
well with the observed quiescent luminosities of NS LMXBs, though some
systems like Cen X-4 (Rutledge et al. 2001a) and KS 1731--260 (Wijnands
et al. 2001; Rutledge et al. 2002b) may be problematic.

Equation (\ref{accdch}) above is general and is not necessarily
restricted to a NS.  It ought, therefore, to apply to J1118 if the
object has a surface crust of normal matter.  The total energy emitted
by J1118 during its accretion outburst of 2000 is estimated to be
$S_{\rm acc} \sim 1.6\times10^{43} ~{\rm erg}$.  This estimate was
computed using the 1.5--12 keV flux observed by the All-Sky Monitor
aboard RXTE over the entire 2000 outburst ($\approx$200 days), and by
approximating the source spectrum as a pure power law with photon
index $\Gamma = 1.78$ extending over the energy range 0.3--120 keV
(McClintock et al. 2001b).  The accretion efficiencies $\eta$ of
Models 1 to 9 described in \S4 and Appendix A vary from 0.18 to 0.67, 
with a
typical value of about 0.3.  Using the latter value, we estimate that
J1118 should, as a result of the 2000 outburst, have a net fluence of
escaping deep crustal heat of order $S_{\rm dch} \sim 8\times10^{40}$
erg.  The duration over which the heat escapes is given by the
diffusion time from the layer where the reactions occur to the
surface.  For lack of a better estimate, we take the time to be
$\sim1-10$ yr, as calculated by Brown et al. (1998) for NSs.  We then
predict a post-outburst quiescent thermal luminosity in J1118 of $\sim
2.5\times10^{32} - 2.5\times10^{33} ~{\rm erg\,s^{-1}}$.  This is
tens to hundreds of times larger than the conservative 99\% confidence
level upper limit we have derived from our observations.

An alternative approach is to consider the time averaged accretion
rate of J1118 (rather than the mass accreted during the last outburst)
and to estimate thereby the mean deep crustal heating rate.  For a
slightly evolved secondary and a binary with an orbital period of 4.1
hr, King, Kolb, \& Burderi (1996) estimate the mass transfer rate from
the secondary to be $\sim 10^{-10} ~{\rm M_\odot yr^{-1}}$.  Assuming
that most of the transferred mass ultimately accretes onto the compact
star, this gives a mean deep crustal heating rate of $9\times10^{33}
~{\rm erg\,s^{-1}}$, which is $\approx 1000$ times larger than our
upper limit on the quiescent thermal flux from J1118.

Both the above calculations indicate that J1118 is anomalously dim in
quiescence compared to straightforward applications of the deep
crustal heating model.  One might be able to contrive a version of the
model that brings the model predictions below the observed luminosity
limit.  Parameters that one could play with are the conductivity of
the crust and the diffusion time from the heating layer to the
surface.  An exploration of these possibilities is beyond the scope of
this paper.  Colpi et al. (2001) invoked enhanced neutrino cooling in
the core of the neutron star in Cen X-4 to explain the relatively low
luminosity in quiescence of that source.  A similar explanation, but
with more extreme parameters, might work for J1118.  This is worth
exploring in more detail.

In our opinion, the most straightforward explanation of the
observations is that J1118 simply does not generate any deep crustal
heat through nuclear reactions.  Since the reactions invoked by Brown
et al. (1998) are unavoidable in normal matter (but see \S 5.3 below),
we suggest that J1118 is a true BH with an event horizon.  The object
has no surface, and hence no opportunity either to undergo deep
crustal nuclear reactions or to radiate the released energy.

\subsection{Thermal Surface Emission from Accretion}

As discussed in \S4.2, the spectra of quiescent NS LMXBs show two
distinct components, a thermal component and a power-law component,
with the former typically having a somewhat larger flux by a factor of
a few.  The most natural explanation for the power-law component is
accretion, since there is no reasonable scenario in which deep crustal
heating or any other energy source inside the NS will lead to
power-law emission.  We consider in this subsection the possibility
that even the thermal component is the result of accretion (e.g.,
Narayan et al. 1997b).  This is a reasonable hypothesis for several
reasons.  First, if the power-law emission arises close to the surface
of the NS, then a good fraction of the radiation will impinge on the
surface of the star and will be thermalized and re-emitted as a soft
thermal component.  Second, several models of accretion onto a NS show
that a fraction of the accretion energy is released below the
photosphere, leading to spectra that consist of both a thermal and a
power-law component (Shapiro \& Salpeter 1975; Turolla et al. 1994;
Zampieri et al. 1995; Zane, Turolla, \& Treves 1998; Deufel, Dullemond
\& Spruit 2001).  Finally, the fact that the thermal component varies
significantly in some sources (Rutledge et al. 2002a) is natural in an
accretion model but very problematic with the deep crustal heating
model.

Since transient BH LMXBs and NS LMXBs are very similar in many
respects, one might expect their luminosities in quiescence to be
comparable.  Indeed, Menou et al. (1999) showed that {\it
Eddington-scaled} quiescent luminosities should be comparable for
systems with similar orbital periods; equivalently, the {\it unscaled}
luminosities of BH LMXBs should be larger by a factor of several (the
ratio of BH mass to NS mass) than the luminosities of NS LMXBs.
Quiescent BH LMXBs are, however, actually seen to be very much fainter
than quiescent NS LMXBs.  This was first pointed out by Narayan et al.
(1997b) and later confirmed in other studies (Menou et al. 1999;
Garcia et al. 2001; MNG03).  Figure \ref{fig:4}\placefigure{fig:4}
shows the latest data, including new results from Tomsick et
al. (2003a,b).  We see that the Eddington-scaled luminosities of BH
LMXBs are orders of magnitude less than those of NS LMXBs with
comparable orbital periods.  Figure \ref{fig:5}\placefigure{fig:5}
shows the same results without using the Eddington scaling.  Even in
this representation, it is clear that quiescent BH LMXBs are very much
fainter than quiescent NS LMXBs, whereas they should be brighter
according to the Menou et al. (1999) scalings.

The large difference between BH LMXBs and NS LMXBs in quiescence is
surprising because the two sets of sources are expected to have
similar mass transfer rates and mass accretion rates (e.g., King et
al. 1996 and the discussion in MNG03).  One potential explanation for
the difference is that BH LMXBs radiate most of their emission in a
thermal component and that this component is much cooler than in NS
LMXBs because of the larger surface area of BH candidates.  The
thermal component in a BH candidate might then be so cool that it is
heavily absorbed by the ISM, leading to a spuriously low estimate of
the luminosity of the system.

The above hypothesis can be tested with J1118 since the source has an
extraordinarily low interstellar column, so that even a very cool
thermal component can, in principle, be detected.  Combining the 99\%
confidence level upper limit on the thermal component of $9.4\times
10^{30} ~{\rm erg\,s^{-1}}$ derived in \S4.1, with the power-law
luminosity of $3.5\times10^{30} ~{\rm erg\,s^{-1}}$ reported in MNG03,
we derive an upper limit of $1.3\times10^{31} ~{\rm erg\,s^{-1}}$ for
the total quiescent luminosity of J1118.  This limit is shown in
Figures \ref{fig:4} and \ref{fig:5} as a filled square.  Even after
allowing for the maximum amount of soft thermal emission that J1118
can possibly have, we see that the quiescent luminosity of this source
is still very much less than that of comparable NS LMXBs.

The large difference between BH LMXBs and NS LMXBs in quiescence finds
a natural explanation if BHs have event horizons.  Narayan et
al. (1997b) proposed that (i) accretion in quiescent LMXBs occurs via
an advection-dominated accretion flow (ADAF; Narayan \& Yi 1994, 1995;
Abramowicz et al. 1995; Narayan, Mahadevan \& Quataert 1998), for
which there is considerable supporting evidence (Narayan, McClintock
\& Yi 1996; Narayan, Barret \& McClintock 1997a; Esin, McClintock \&
Narayan 1997; Esin et al. 1998, 2001), and (ii) that BH LMXBs are
unusually dim because they advect most of the accretion energy into
the BH through the event horizon, in contrast to NS LMXBs which
radiate the advected energy from the surface of the NS.  According to
this model, the difference in luminosity between quiescent BH LMXBs
and NS LMXBs is a direct consequence of the fact that BH candidates
have event horizons whereas NSs have surfaces.  

Several authors have proposed astrophysically motivated
counter-explanations for the data shown in Figures \ref{fig:4} and
\ref{fig:5} (Bildsten \& Rutledge 2000; Stella et al. 2000; Nayakshin
\& Svensson 2001; Fender, Gallo \& Jonker 2003), but many of these
proposals have encountered difficulties as a result of later
observations (see Narayan et al. 2002 and Kong et al. 2002).

\subsection{Exotic Physics}

The previous two subsections show that, within the realm of
astrophysically reasonable explanations, it is hard to understand (i)
the lack of a thermal component in the spectrum of J1118, and more
generally, (ii) the low luminosity in quiescence of the source.  The
most reasonable explanation is that the object is a black hole.  We
also note the argument of Narayan \& Heyl (2002), according to which
the absence of Type~I X-ray burst from J1118 argues against the
presence of a surface.  Here we investigate models based on exotic
physics to see if any of these are consistent with the observations.
The literature on exotic models is vast, and we are able to deal with
only a small fraction of it here.

Glendenning (2000), Haensel (2004) and Weber (2004) discuss several
exotic alternatives to the traditional neutron star, including
relatively minor modifications such as a neutron star with a pion or
kaon core as well as other more extreme possibilities such as hyperon
stars, quark-hybrid stars, strange stars, etc.  Most of these
alternatives give viable models only for stellar masses $\lesssim 2-3
M_\odot$ and are not relevant for J1118 or other black hole
candidates, which are more massive.  Nevertheless, one might wish to
consider whether such models would be consistent with the data on
J1118 if they could somehow survive at a mass $\sim8M_\odot$.

Bodmer (1971) and Witten (1984) suggested that the true ground state
of the strong interaction may not be $^{56}$Fe, but a state of
deconfined quark matter in a color superconducting state.  If this is
the case, it would be possible to have a ``strange star'' that is made
largely of such quark matter (e.g., Dey et al. 1998).  Such stars tend
to be more compact than neutron stars (though still respecting the
Buchdahl limit, $R<9/8 R_S$).  The maximum mass of these models is
$<3M_\odot$ (Glendenning 2000; Haensel 2004; Prasad \& Bhalerao 2004),
so they are usually considered as an alternative to neutron stars and
not as a model of a black hole candidate like J1118.  In addition,
when one of these stars accretes gas, it develops a surface crust of
normal matter (e.g., Usov 1997; Zdunik 2002; Weber 2004) that floats
above the quark substrate because of a repulsive Coulomb barrier.  The
crust extends down to a density of order the neutron drip point ($\sim
10^{12} ~{\rm g\,cm^{-3}}$), and is indistinguishable from the crust
of a normal neutron star.  Consequently, the star will undergo deep
crustal heating reactions (\S 5.1) and Type I X-ray bursts (Narayan \&
Heyl 2002), and will release accretion energy in the form of X-rays
(\S5.2), just as in the standard model.  (See Sinha et al. 2002 for a
discussion of possible burst characteristics of strange quark stars.)
Therefore, even if the model could be extended to the mass range of
black hole candidates (which is unlikely, see Glendenning 2000; Weber
2004), it could still be ruled out for J1118.

Alford, Rajagopal \& Wilczek (1999) showed that at asymptotically
large densities, a color-flavor locked (CFL) state of up, down and
strange quarks is favored over other forms of color superconducting
quark matter.  Compact stars made of CFL matter have been considered
by some authors.  The maximum mass of the star appears to be about
$4.5 M_\odot$ (Horvath \& Lugones 2004), which is too low for most
black hole candidates.  If the CFL phase is not the global energy
minimum state at zero pressure, then a CFL star would consist of CFL
matter in the core and normal matter in the crust (e.g., Alford et
al. 2001).  Such a model can be ruled out by the arguments presented
in \S\S 5.1, 5.2.  However, if the CFL matter is the true ground state
at zero pressure, then it is possible to have a bare CFL star.  

In contrast to the strange star discussed earlier, a bare CFL star has
no electrical charge and therefore does not support a crust
electrostatically.  Thus, one could visualize a situation in which
accreting gas is immediately converted to CFL matter on contact with
the surface of the star.  Since no baryons or nuclei survive, there
will be no deep crustal heat generation (or Type I X-ray bursts,
Abramowicz, Kluzniak \& Lasota 2002).  Moreover, CFL matter is a very
poor emitter of electromagnetic radiation at keV temperatures
(Jaikumar, Prakash \& Sch\"afer 2002), so the star will be very dim.
An object like this would be hard to rule out by means of observations
(Horvath \& Lugones 2004).  Another variant is a CFL star with a crust
made up of a two color-flavor superconductor (2SC) or some other phase
of quark matter (Weber 2004).  The second phase does have electric
charge (including free electrons) and it is expected to be an
efficient radiator of electromagnetic radiation (Page \& Usov 2002).
Yet other forms of quark matter and their possible relevance to
compact stars are reviewed by Weber (2004), but we do not discuss them
here.

In summary, the situation with respect to models based on the CFL
phase is fairly uncertain.  Certain types of CFL stars --- the bare
variety with CFL matter extending out to the surface --- are
indistinguishable from black holes, at least via electromagnetic
signals.  But more study of the CFL phase is needed before one can
decide if these models are viable.  In particular, it is important to
determine if the CFL phase is the ground state of quark matter at zero
pressure, and even if it is, whether models based on it can exist at
masses approaching $10M_\odot$.

Some theories of the strong interaction allow nucleons to be confined
at densities below nuclear density.  Compact stars made of this
so-called Q-matter have been considered by Bahcall, Lynn \& Selipsky
(1990) and Miller, Shahbaz \& Nolan (1998).  A feature of Q-stars is
that their maximum mass can be quite high, in principle even above
$100 M_\odot$ if the Q-phase can exist at sufficiently low density.
These models are thus relevant for black hole candidates.  There is
very little discussion in the literature on what happens when gas
accretes on a Q-star.  Since Q-matter is made of normal baryons and
electrons, accreting gas should form an electrostatically supported
crust of normal matter, just as in the case of the (non-CFL) strange
star discussed earlier.  If this is the case, then a Q-star will have
deep crustal reactions and Type I X-ray bursts, and will release
accretion energy in the form of electromagnetic radiation.  The model
can then be ruled out by the observations.

Moving on to more exotic possibilities, one could consider a
two-component model of a compact star in which one of the components
is ordinary gas and the other is a non-interacting form of dark
matter.  The dark matter, which dominates the mass, may be either
fermionic or bosonic.  Such models have been studied in the past (Lee
\& Pang 1987; Zhang 1988; Henriques, Liddle \& Moorhouse 1989; Jin \&
Zhang 1989; Yuan, Narayan \& Rees 2004).  In a related class of
models, the dark component may consist of shadow or mirror matter
living in a different sector (Khlopov et al. 1989).  A feature of all
these models is that the gas component has radii, surface gravities
and surface redshifts similar to those of a neutron star (Yuan et
al. 2004).  Thus, the objects are expected to behave like neutron
stars when they accrete gas.  Specifically, they ought to release a
similar amount of heat through deep crustal reactions and should be as
bright as neutron star systems at the same mass accretion rate.  As
the discussion in \S\S 5.1, 5.2 indicates, J1118 behaves quite
differently from a neutron star and thus is not consistent with any of
these models.

Chapline et al. (2001) have proposed that the event horizon of a
classical black hole corresponds to a quantum phase transition with
properties similar to the critical point of a Bose fluid.  In this
interpretation, there is no true event horizon.  Therefore, a black
hole will radiate whatever energy accretes on it, and it is not
expected that black hole X-ray binaries in quiescence should be
anomalously dim compared to neutron star systems.  However, the
spectrum of the emitted radiation might be quite different from a
simple blackbody (Barbieri, Chapline \& Santiago 2003).  It would be
of interest to calculate the spectrum of a stellar-mass quantum black
hole in a quiescent X-ray binary to check whether the model is
consistent with the observations shown in Figures 4 and 5.

Robertson and Leiter (2002, 2003) have used a new class of solutions
of the Einstein field equations of general relativity to describe
black hole candidates as magnetospheric, eternally-collapsing objects
(MECO) that have a surface redshift $z \sim 10^8$.  The magnetic field
is assumed to be anchored and corotating with the central object.  The
predicted magnetic spindown luminosity and spectrum of the power-law
component in quiescence is consistent with the observed values for
J1118 (\S1) for an intrinsic magnetic moment of $\sim 3.4 \times
10^{29}$ G cm${^3}$, assuming that the dipole is aligned along the
spin axis and spinning slowly at $\sim 8$~Hz (D. Leiter, private
communication).  Our limits on the temperature and luminosity of the
thermal component of J1118 (\S4.1) can be accommodated if the surface
redshift of the object exceeds $1.39 \times 10^8$ (eqn. 17 in
Robertson and Leiter 2003).

Finally, Abramowicz et al. (2002) invoked the gravitational condensate
star model (or gravastar model) of Mazur \& Mottola (2002) and argued
that there is no way to distinguish such an object, which has a
surface at a radius $R$ only slightly greater than $R_S$, from a
genuine black hole with an event horizon.  They make this claim based
on the surface redshift, which is very large (because $R$ is close to
$R_S$).  However, if the source is in steady state and the luminosity
is the result of accretion, then, regardless of the surface redshift,
the binding energy of the accreting gas has to be radiated to
infinity.  In fact, the more compact the star, the larger the amount
of binding energy released per unit accreted rest mass, and the larger
the luminosity at infinity for a given mass accretion rate!  Thus, the
Abramowicz et al. argument is not valid for our problem; that is, the
luminosity at infinity does {\it not} decrease as the surface redshift
increases.  However, we should note that the luminosity might come out
in a form other than electromagnetic radiation, e.g., neutrinos or
some kind of exotic particles (since the radiating gas can be very
hot).  Clearly, such models cannot be ruled out by the present work.

We should note that the gravitational condensate star is able to
violate the Buchdahl limit, $R \ge 9/8 R_S$, only because it invokes a
negative-pressure interior (filled with vacuum energy, $p=-\rho$).  It
also has an anomalously low entropy (Abramowicz et al. 2002).  These
features are unpalatable to many physicists.

\section{Conclusions}

We have examined the possibility that the dynamical black-hole
candidate J1118 possesses a material surface rather than an event
horizon.  Either accretion onto such a surface or deep crustal heating
would be expected to produce a quiescent thermal component of emission
like those commonly observed for neutron stars.  We have fitted our
Chandra spectrum of J1118 to a model consisting of a fixed power-law
component plus an atmospheric thermal component with variable
temperature, $\Tinf$.  The spectral fits were repeated for a series of
nine atmospheric models with radii ranging from the minimum allowable,
$9/8\, \Rs$, to a maximum of $2.8\, \Rs$.  For the most compact of
these models, which lie within their own photon spheres, the
self-irradiation of the atmospheres was taken into account.  No
emission in excess of a simple power-law component was detected in
J1118, and very strong upper limits were set on the presence of a
thermal source: $k\Tinf <$ 0.011 keV and $L_{\infty,\rm th} < 9.4
\times 10^{30}$ \lum~(99\% confidence level).

If one assumes that the hypothetical crust of J1118 is composed of
normal nuclear matter, then this stringent limit on a thermal
component of luminosity is hard to reconcile with the theory of deep
crustal heating and the observed fluence of J1118 during its outburst
in 2000: The predicted quiescent luminosity exceeds the above limit on
$L_{\infty,\rm th}$ by a factor of $\gtrsim 25$.  Possibly a contrived
model of deep crustal heating and/or an extreme model of neutrino
cooling of the core could explain this difference.  On the other hand,
if J1118 possesses a material surface and accretion powers the thermal
emission seen from NSs, then one expects J1118 to have a luminosity at
least as great as that of an average NS, whereas its {\it total}
luminosity in Eddington-scaled units is about 100 times less than the
luminosity of a typical NS and fully 10 times less than the luminosity
of even SAX J1808.4-3658 (Fig. 4).

The above limit on thermal emission, in combination with the observed
power-law emission, yields a very tight limit on the {\it total}
quiescent luminosity of $1.3 \times 10^{31}$ \lum, which is far below
the luminosities observed for NSs.  Because of the high transparency
of the ISM, our results rule out the possibility that the total
luminosity of J1118 could be augmented significantly by any ultrasoft
component of emission.  Thus J1118 -- and by inference the other
dynamical BH candidates -- are truly faint relative to NSs
(Figs. \ref{fig:4} and \ref{fig:5}).

In summary, a sensitive search has failed to detect any thermal
emission from a hypothetical surface surrounding J1118, although NSs
very commonly show such surface emission due to either deep crustal
heating or accretion.  Our sensitivity to a thermal component of
emission from J1118 is much greater than the emission predicted by the
theory of deep crustal heating, assuming that J1118 has a material
surface analogous to that of NSs.  Likewise, there is no evidence that
accretion is occurring in quiescence onto the surface of J1118, which
is the mechanism often invoked to explain the far greater thermal
luminosities of NSs.  The simplest explanation for the absence of any
thermal emission is that J1118 lacks a material surface and possesses
an event horizon.

Finally, our limits on thermal emission from J1118 rule out the
possibility that there is a heretofore unseen and appreciable soft
component of luminosity.  This result implies that the dynamical BH
candidates are truly faint relative to NSs and underscores our
original argument that these compact objects have event horizons and
are therefore genuine black holes (Narayan et al. 1997b).  As
discussed in \S 5.3, however, we cannot at this time rule out certain
very exotic alternatives.

\appendix

\section{Atmospheric Models}

The emergent spectra of the compact objects specified in the previous
section are found by modeling their atmospheres.  Our approach is
similar to the modeling done by Zavlin et al. (1996), except for
certain modifications due to self-irradiation of the atmosphere when
the surface of the object is inside its own photon sphere.

The atmospheric models used here incorporate relatively simple
physics, in keeping with the purpose of the paper to provide bounds,
rather than detailed spectral comparisons.  The major assumptions are:

\begin{enumerate}

\item Negligible magnetic field.

\item 
Spherical object with static, plane-parallel atmosphere in radiative
equilibrium.

\item 
Ideal equation of state for pure hydrogen with complete ionization.

\item 
Opacity due to free-free absorption plus Thomson scattering in
the unpolarized, iso\-tropic approximation.  

\end{enumerate}

A few comments on the above assumptions: Negligible magnetic field $B$
in this context implies $B \ll 10^8$--$10^{10}$ G (Zavlin et al. 1996). 
The scale heights in the atmosphere are much less than the radius, so
the plane-parallel assumption is a good one.  The assumption of pure
hydrogen is appropriate for a slowly accreting object with sufficient
time for gravitational settling.  Within the zones where ideal
behavior is valid, complete ionization is a good assumption at the
highest effective temperatures ($\sim 10^6$ K) considered, but will be
less good at the lowest ($\sim 10^5$ K).  The complete ionization
assumption implies that bound-free and bound-bound opacities need not
be included.  As indicated above, this is less valid for the lowest
temperatures considered.  However, even there the bound-free and
bound-bound features are still well below the peak of the spectrum,
and would probably have only a minor effect on our conclusions.

We shall now describe the relevant general
relativistic effects, followed by our results for the atmospheric
modeling.

\subsection{General Relativistic Effects}  

Schwarzschild geometry exactly
describes the region outside the spherical star, and is also
applicable to a high degree of approximation within the thin
atmosphere.  Here we review aspects of Schwarzschild geometry relevant
to the present discussion only briefly, referring to discussions in
standard texts, especially Misner, Thorne \& Wheeler (1973, hereafter 
MTW).
Using standard notation (MTW, p.\ 655), the metric in Schwarzschild
coordinates is,
\be
  ds^2 = -(1-\Rs/r)\,dt^2 + (1-\Rs/r)^{-1}\,dr^2 + r^2\, 
   ( d\theta^2 + \sin^2\theta \, d\phi^2);~~\Rs= 2GM/c^2.    \eqlab{gbr1}
\ee
It follows that the gravitational redshift and time dilation are given
in terms of the redshift factor 
\be 
1+z=(1-\Rs/r)^{-1/2}.
\eqlab{gbr2} \ee In particular, a photon of frequency $\nu$ at the
surface radius $r$ will be observed at frequency
$\nu_\obs=(1+z)^{-1}\nu$ at large radius $r_\obs$, $r_\obs \gg \Rs$.
Likewise a time interval $dt$ at the surface will appear as
$dt_\obs=(1+z)\,dt$ at $r_\obs$.  One notes that the product
$dt\,d\nu=dt_\obs\,d\nu_\obs$ is therefore an invariant.

It is important to determine how the photon flux observed at a large
radius $r_\obs$ is related to the flux at the surface radius $r$.  
[In this Appendix, $r$ is used to denote both a general radius and also
the radius of the object, called $R$ in the main body of the paper.]
One notes that in steady state the same number of photons $dN$ flowing
out {}from the surface in a time interval $dt$ will pass through the
spherical surface at $r_\obs$ in time interval $dt_\obs$.  Since the
area of a spherical surface in Schwarzschild geometry is simply
proportional to the square of the Schwarzschild radial coordinate,
this implies that the photon flux expressed as number of photons per
area per time per frequency interval transforms in the following way:
\be
   \left( {dN \over dA\,dt\,d\nu} \right)_\obs
  = {r^2 \over r_\obs^2} \left( {dN \over dA\,dt\,d\nu} \right)_{\rm 
surface}.
  \eqlab{gbr3}
\ee

Often it is more convenient to use a flux based on photon
energy, $F=dE/dA\,dt\,d\nu$.  Since $dE=h\nu\,dN$, this implies the
transformation of flux,
\be
   F(r_\obs,\nu_\obs) =(1+z)^{-1}{r^2 \over r_\obs^2} 
F[r,(1+z)\nu_\obs].   \eqlab{gbr4}
\ee
The surface flux $F(r,\nu)$ is found by atmospheric modeling.  
The observed flux $F(r_\obs,\nu_\obs)$ can then be determined using 
equation
(\eqref{gbr4}).  The transformations of other quantities related to
flux can be found in similar fashion.

As an instructive example, consider the special case where the surface
emits as a blackbody of temperature $\Teff$.  The surface flux is then
given in terms of the Planck function $B(\nu,T)$,
\be
   F(r,\nu)=\pi B(\nu,\Teff) 
     = \pi {2h\nu^3/c^2 \over \exp(h\nu/k\Teff)-1}.  \eqlab{gbr5}
\ee
Substitution of this into Eq.\ (\eqref{gbr4}) gives,
\be
   F(r_\obs,\nu_\obs)= (1+z)^2 {r^2 \over r_\obs^2} \pi 
B(\nu_\obs,\Tinf),
    \eqlab{gbr6}
\ee
where,
\be
       \Tinf = { \Teff \over 1+z}.  \eqlab{gbr7}
\ee
Although this is a very special case, the form of Eq.\ (\eqref{gbr6})
suggests that in general $\Tinf$ may better characterize the shape of
the observed spectrum than $\Teff$.  Accordingly, we shall use $\Tinf$
to parameterize our models.

So far we have considered the {\em global} consequences
of general relativity, those determining the relationship
between local quantities at the surface of the object with observed
quantities at large distance.  We now shall consider what {\em local}
consequences there might be on the structure of the atmosphere itself.  

To the extent that the atmosphere depends only on local variables,
such as effective temperature and gravity, one can consider the
atmospheric modeling problem as identical to the ``standard'' one by
moving into the local proper frame of the atmosphere.  Since the
chemical composition is fixed here (pure ionized hydrogen), we shall
regard the two local quantities $\Teff$ and $g$ as constituting the
complete parameterization of the atmosphere.  If the mass and radius
of the object are given, then $g$ is given in Schwarzschild geometry
by the corrected Newtonian formula, 
\be 
g=(1+z) {GM \over r^2},
\eqlab{bgr8} 
\ee 
for an object of mass $M$ and radius $r$ (Zeldovich \& Novikov 1971,
Eq.\ (3.2.3), p.\ 87).  The above simple local picture is valid as
long as the radius $r$ of the atmosphere lies outside the photon orbit
corresponding to mass $M$, i.e., the radius satisfies $r>(3/2)\Rs$.
In this case all surface rays emergent in the outward hemisphere
escape to infinity, although they may suffer gravitational bending.
Conversely, all rays in the inward hemisphere come from infinity and
have zero intensity.  In the language of stellar atmospheres, the
boundary condition on the intensities at the surface is
\be 
I(r,\nu,-\mu) = 0, \qquad 0 \le \mu \le 1, \eqlab{gbr9} 
\ee 
where $\mu=\cos\theta$ and $\theta$ is the angle between the ray and
the outward normal.  This is the standard boundary condition, so the
construction of the atmosphere is ``standard'' in every sense.

However, when $r<(3/2)\Rs$ the surface lies inside the {\em photon 
sphere}.
In this case only those rays within a certain critical cone about the
outward normal
will escape, that is, for
$\theta < \thetac$, where [MTW, p.\ 675]
\be
     \sin \thetac = {3^{3/2} \over 2} \left( 1-{\Rs \over r} 
\right)^{1/2}
                    {\Rs \over r}.  \eqlab{gbr10}
\ee
Rays outside this critical cone, $\theta > \thetac$, will be 
gravitationally bent enough to cause them
to return to the surface, at the same angle, but now with respect to 
the inward normal.  Thus a portion of the incoming radiation
will have non-zero intensity, in fact the same intensity as the
emergent rays for the corresponding angle. The boundary condition then
becomes,
\be
   I(r,\nu,-\mu) = \cases{0,   &\qquad$\muc \le \mu \le 1$,\cr
                 I(r,\nu,\mu), & \qquad $0 \le \mu \le \muc$}  
\eqlab{gbr11}
\ee
where
\be
         \muc = \cos\thetac.   \eqlab{gbr12}
\ee
It is seen that the rays outside the critical cone are in effect
specularly reflected (angle of incidence equals angle of reflection).
This {\em self-irradiation} is a nonlocal effect of general relativity
that does not have any analogy in the ``standard'' atmosphere problem.

The most compact object to be considered here is one at the Buchdahl
limit $(9/8)\Rs$, where the critical cosine is
$\muc=(11/27)^{1/2}=0.6383\ldots$.  Thus, there is potentially a 
reduction in the available solid angle for escaping rays by a factor
of order two or more.  

Given the unusual nature of the intensity field at the surface in 
the self-irradiation case, it is worthwhile to re-examine the definition
of surface flux.  We start with the formula for flux for
an object with spherical symmetry, 
\be
       F(r,\nu) = 2\pi \int_{-1}^{1} I(r,\nu,\mu)\mu\,d\mu.  
\eqlab{gbr13}
\ee
The effective temperature $\Teff$ is defined using the total surface 
flux,
which is found by integrating $F(r,\nu)$ over frequencies,
\be
        \sigma \Teff^4 = \int_0^{\infty} F(r,\nu)\, d\nu,  \eqlab{gbr14}
\ee
where $\sigma$ is the Stefan-Boltzmann constant.  Equations 
(\eqref{gbr13}) and (\eqref{gbr14}) apply to all cases.  

We shall now show how (\eqref{gbr13}) simplifies in the standard case
and in the self-irradiation case.  In the standard case, $r>(3/2)\Rs$,
the intensity vanishes for $\mu <0$ (cf.\ Eq.\ [\eqref{gbr9}]), so
that equation (\eqref{gbr13}) can be written as an integral over the
outward hemisphere alone,
\be
 F(r,\nu) = 2\pi \int_{0}^{1} I(r,\nu,\mu)\mu\,d\mu,  \qquad r>(3/2)\Rs. 
     \eqlab{gbr15}
\ee

In the self-irradiation case, $r<(3/2)\Rs$, we break the range of
integration in Eq.\ (\eqref{gbr13}) into four intervals:
($-1$,$-\muc$), ($-\muc$,$0$), ($0$,$\muc$), and ($\muc$,$1$).  From
equation (\eqref{gbr11}), it is seen that the integral over the first
interval is zero, while the integrals over the second and third cancel
due to the reflection condition and the odd factor $\mu$.  Therefore,
the flux can be written as an integral over the fourth interval alone,
\be
 F(r,\nu) = 2\pi \int_{\muc}^{1} I(r,\nu,\mu)\mu\,d\mu,  
    \qquad r<(3/2)\Rs, \eqlab{gbr16}
\ee
In either case, (\eqref{gbr15}) or (\eqref{gbr16}), the integration
involves only escaping rays, as one would expect intuitively.  Here
we adopt the convention that $\muc=0$ when $r>(3/2)\Rs$, making equation
(\eqref{gbr16}) generally valid.

\subsection{Calculation of the Atmospheric Models}

A code was written to solve for the emergent spectrum of the
atmosphere.  This code uses fairly standard techniques of
discretization in depth, frequency, and angle, along with an iterative
linearization technique for the temperature correction.  The
linearization is only ``partial,'' in that the temperature dependences
of the opacities on the temperature were not linearized, but rather
updated after each iteration.  The convergence rate was nevertheless
quite adequate.

A novel feature of the code is its ability to treat self-irradiation,
which required the incorporation of the modified surface boundary
condition (\eqref{gbr11}).  This also involved using separate
quadrature for the angular ranges inside and outside the critical
cone.  It was found that, per hemisphere, 3 angles within the cone and
2 outside gave adequate results.

In order to check the code, we computed a standard case (no
self-irradiation) with parameters $\log\Teff=5.9$ and $g=2.43\times
10^{14}$ cm s$^{-2}$, a case considered by Zavlin et al. (1996).  Our
emergent intensities at values of $\mu=0.1$, $0.4$, $0.7$, and $1.0$
were compared to theirs (the left side of their Figure 4) with
virtually perfect agreement.

The emergent flux vs.\ frequency of standard models for $\log \Teff =$
6.0, 5.5, and 5.0 are plotted in Figure
\ref{fig:6}.\placefigure{fig:6} These models are for $\log g=14.1$,
but the gravity dependence is not strong, and these curves apply
fairly well to the other gravities considered here.  Note in this
figure the abscissa is $\log (\nu/\Teff)$ and the ordinate is $\log
(F_\nu /\Teff^3)$.  With these rescalings, the various ``standard''
models lie close to a universal curve, which we have fitted to the
form, 
\be 
{F_\nu \over \Teff^3} = 2.23 \times 10^{-17} q^{2.5} \exp
\left[ -q^{0.55} \right], \eqlab{gbr17} 
\ee where, 
\be 
q=1.62 \times 10^{-10} {\nu \over \Teff}, \eqlab{gbr18} 
\ee 
which does not contain $g$.  All quantities are in cgs units.  We give
this approximation for those who may need a quick, convenient way to
estimate such spectra without doing the detailed atmospheric modeling.
However, for the main calculations of this paper, we computed each
individual model using the full code.

The primary effect of self-irradiation is to limit the available solid
angle for escaping photons at the surface, which has the secondary
effect of forcing the temperatures within the atmosphere to rise for
the same effective temperature, thereby acting very much like the well
known backwarming effect for line blanketing in stellar atmosphere
theory (see, e.g., Mihalas 1978).  This is shown in Figure
\ref{fig:7},\placefigure{fig:7} where temperature vs.\ pressure
(depth) is given for models each with $\log g = 14.654$ and $\log
\Teff = 6.0$, but for values of $\muc=0$, 0.1094, 0.3522, and 0.6383.
The first of these corresponds to a ``standard'' model, while the last
corresponds to one with maximal self-irradiation (at the Buchdahl
limit).  The self-irradiated models are hotter than the standard one
at all depths, but especially near the surface.

As shown in Figure \ref{fig:8},\placefigure{fig:8} the main effect of
self-irradiation is to soften the spectrum.  There are two reasons for
this.  First, the escape cone for emergent radiation is narrowed,
which lowers the flux at both low and high frequencies.  Second, the
temperatures near the surface affect mostly the lower frequencies
where the free-free opacities are largest, while the temperatures at
large depth affect mostly the higher frequencies.  As we have seen,
the surface temperature is raised more than those at large depth,
which favors the lower frequencies.  The net result of these two
effects is to yield the softened spectra seen in Figure \ref{fig:8},
but to keep the same total flux, as it must for the same effective
temperature.

It is perhaps worthwhile to point out that the softening of the
spectrum here relies strongly on the substantial frequency dependence
of the free-free opacity, as indicated in the previous paragraph.  In
fact, we have done analogous calculations for a frequency independent
opacity (gray opacity), and have found that in that case the spectrum
actually {\em hardens}.  

\acknowledgments

We thank Keith Arnaud for assistance in implementing table models in
XSPEC and for other advice, Craig Heinke and Peter Jonker for helpful
discussions on quiescent neutron stars, Harvey Tananbaum for comments
on the manuscript, Mike Garcia for help with updating Figures
\ref{fig:4} and \ref{fig:5}, and an anonymous referee for constructive
and stimulating comments.  This work has made use of the information
and tools available at the HEASARC Web site, operated by GSFC for
NASA, and was supported in part by NASA grants NAG5-9930 and
NAG5-10780, and NSF grant AST 0307433.

\newpage

\newpage \figcaption[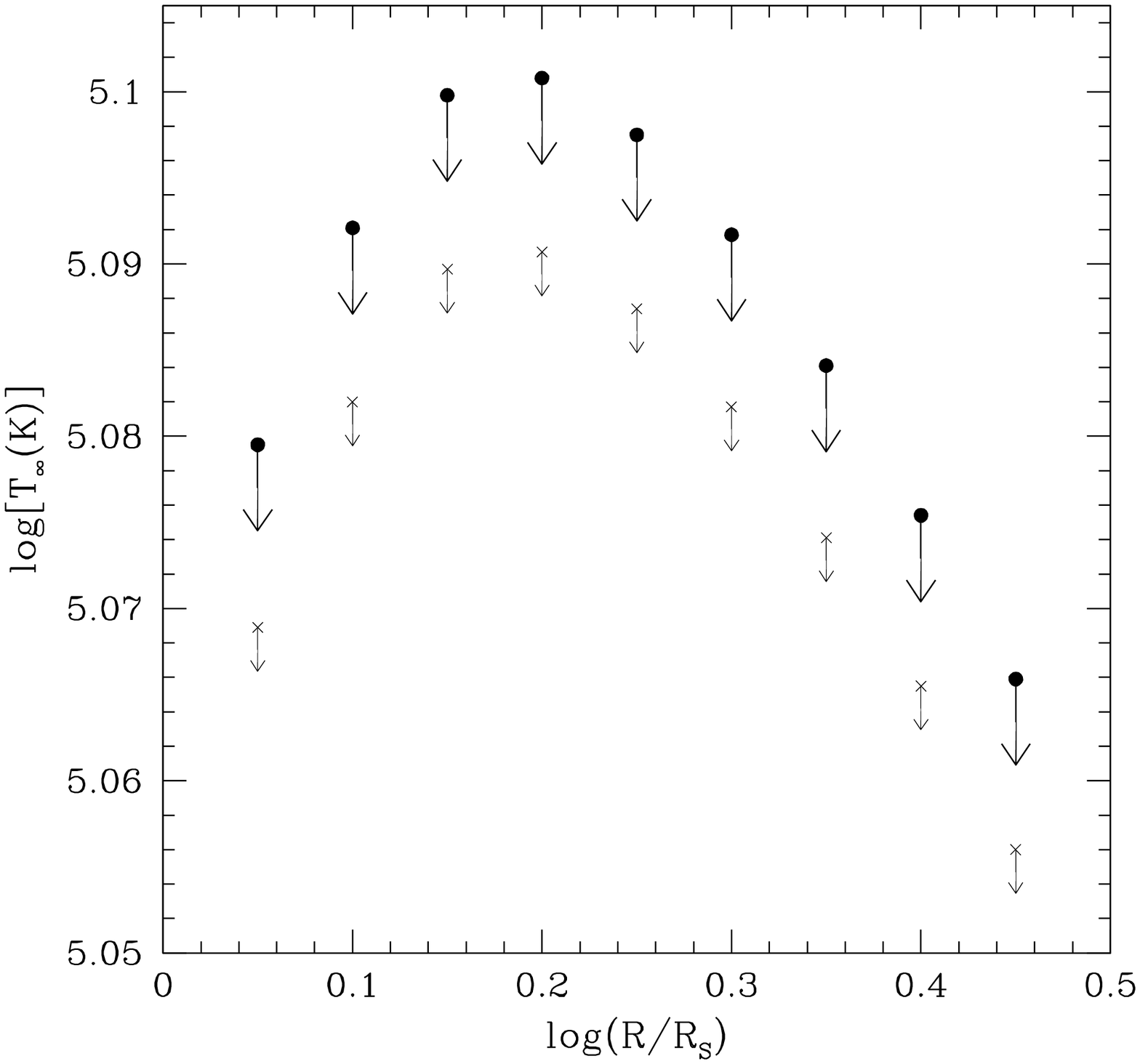] {Upper limits on the temperature of a 
thermal
component of emission as observed at infinity for the nine models
discussed in the text.  The filled circles correspond to the limits
summarized in column 7 of Table 1 and to an adopted maximum
column density of $N_{\rm H}~= 1.6~\times~10^{20}$~cm$^{-2}$ (\S2);
these are the temperature limits considered throughout this work.
More stringent limits that correspond to the most probable column
density of $N_{\rm H}~= 1.2~\times~10^{20}$~cm$^{-2}$ (\S2) are
indicated by the cross symbol.\label{fig:1}}

\figcaption[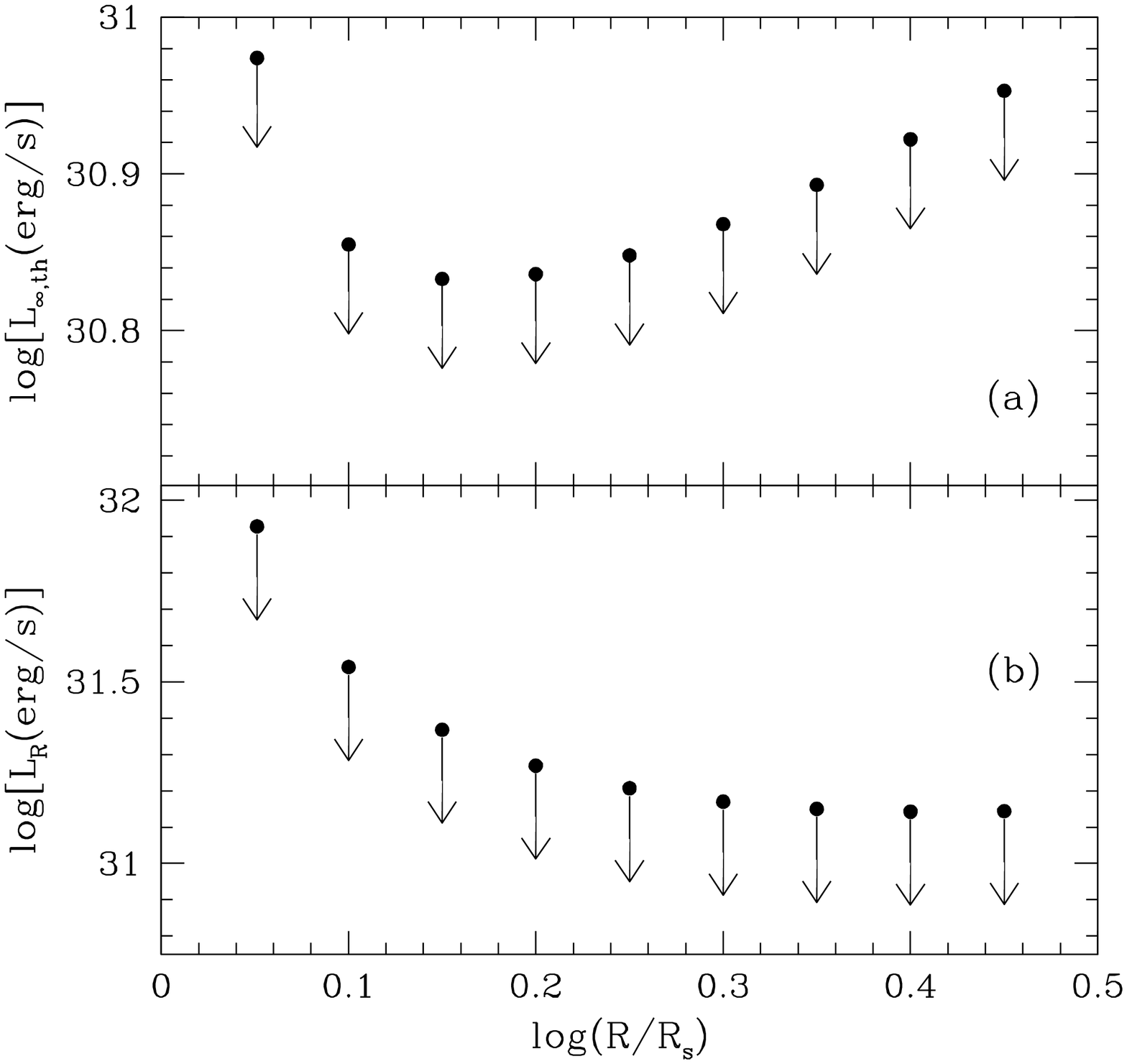] {(a) Upper limits on the luminosity at infinity,
where $L_{\infty,\rm th} = 4 \pi R^{2} \sigma T_{\rm eff}^{4} / (1 +
z)^2$.  The limits assume the maximum adopted column density of
$N_{\rm H}~= 1.6~\times~10^{20}$~cm$^{-2}$ and correspond to the 99\%
level of confidence.  (b) Upper limits on the luminosity at the
surface of the star.\label{fig:2}}

\figcaption[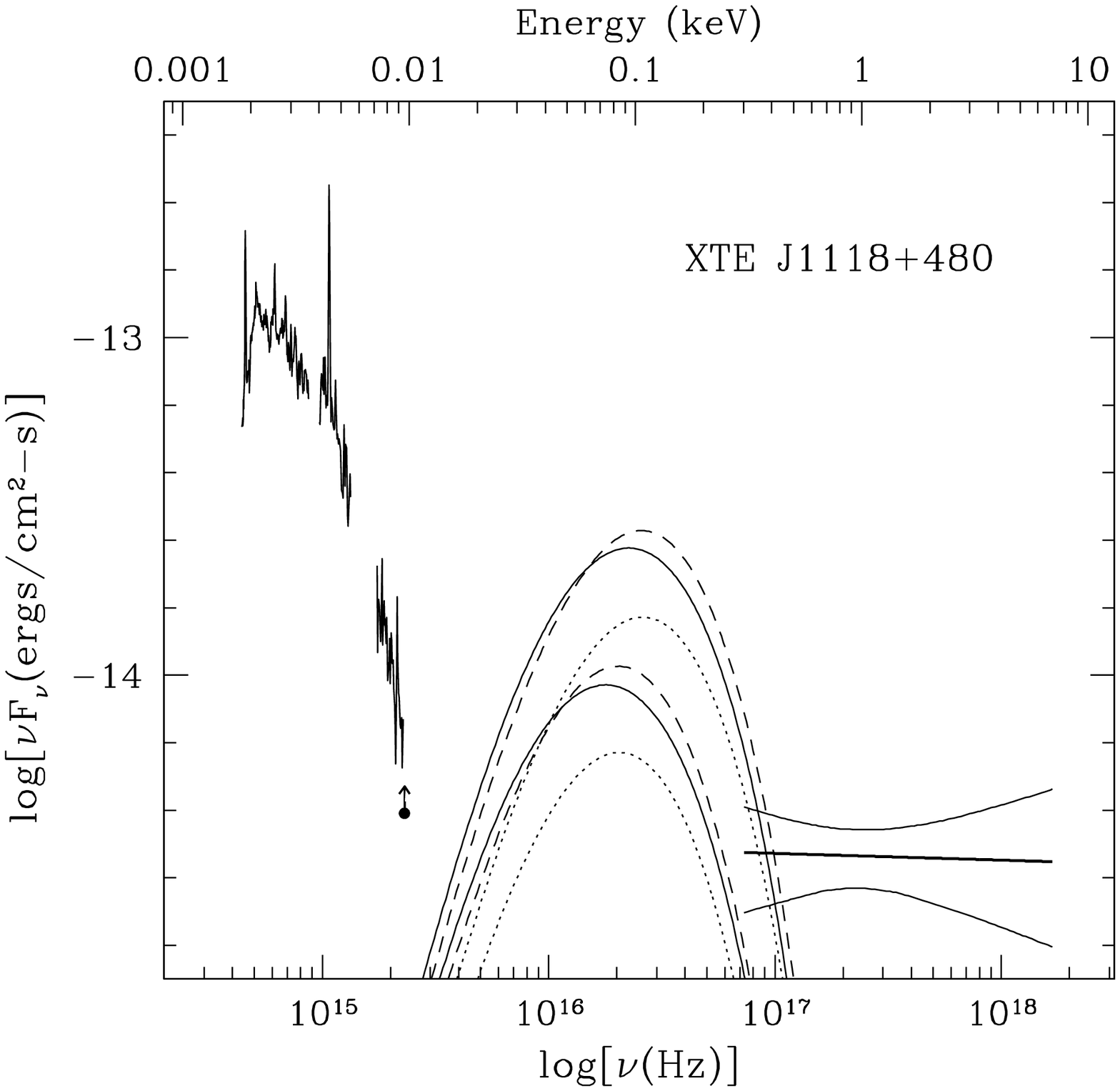] {A comparison of six atmospheric models for
compact stars to a multiwavelength spectrum of J1118, which was
published previously (MNG03).  The three models with the largest peak
fluxes correspond to log$T_{\infty} = 5.15$, and the three
less-luminous models correspond to log$T_{\infty} = 5.05$.  The six
models cover the full range of compactnesses considered in the text
and in Table 1: solid lines, Model 1 (log$R/Rs = 9/8$); dotted lines,
Model 4 (log$R/Rs = 0.20$); and dashed lines, Model 9 (log$R/Rs =
0.45$).  The heavy horizontal line on the right is the best-fit model
X-ray spectrum, and the flanking lines define the 90\% confidence
error box.  The UV/optical spectrum is shown on the left.  Most
relevant is the STIS FUV spectrum centered at $\sim 2 \times 10^{15}$
Hz; these data were obtained in a 14 ks observation with STIS using
the G140L grating.  A firm limit on the sensitivity of this
observation is indicated by the upward arrow.  This limit is centered
at the peak response of STIS and corresponds to the flux at 1300\AA~in
a 100\AA~band; it was computed using HST's web-based Exposure Time
Calculator and assumes average background levels.  HST would fail to
detect a source at this limit (i.e., signal-to-noise ratio $<$
2).\label{fig:3}}

\figcaption[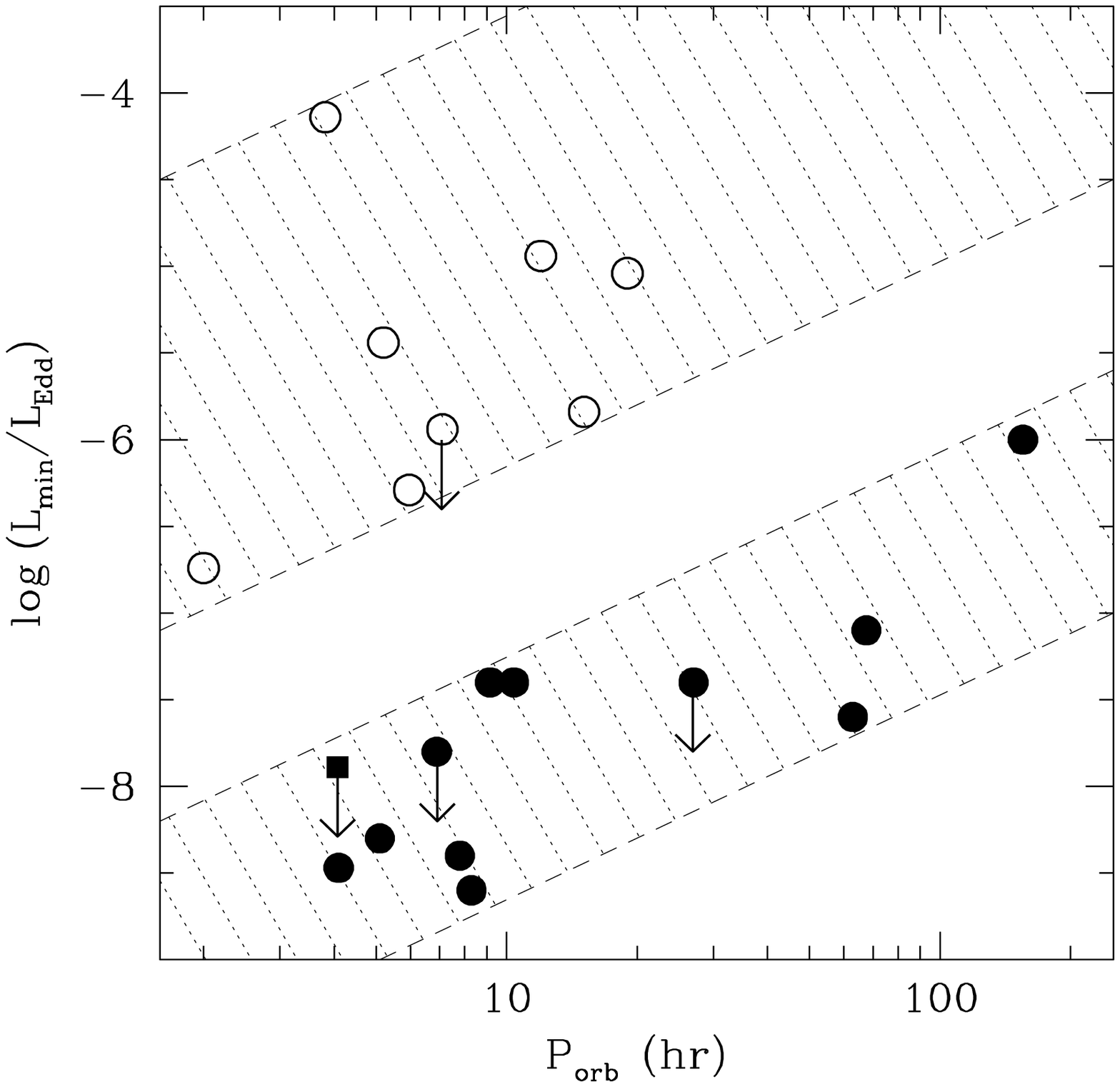] {Eddington-scaled luminosities of BH LMXBs (filled
circles) and NS LMXBs (open circles) vs.\ the orbital period for the
energy range 0.5--10 keV.  The filled square shows the upper limit on
the total flux of J1118 discussed in \S5.2.  Only the lowest quiescent
detections or {\it Chandra/XMM} upper limits are shown.  The diagonal
hatched areas delineate the regions occupied by the two classes of
sources and indicate the dependence of luminosity on orbital period.
Note that BH LMXBs are on average nearly three orders of magnitude
fainter than NS LMXBs with similar orbital periods.\label{fig:4}}

\figcaption[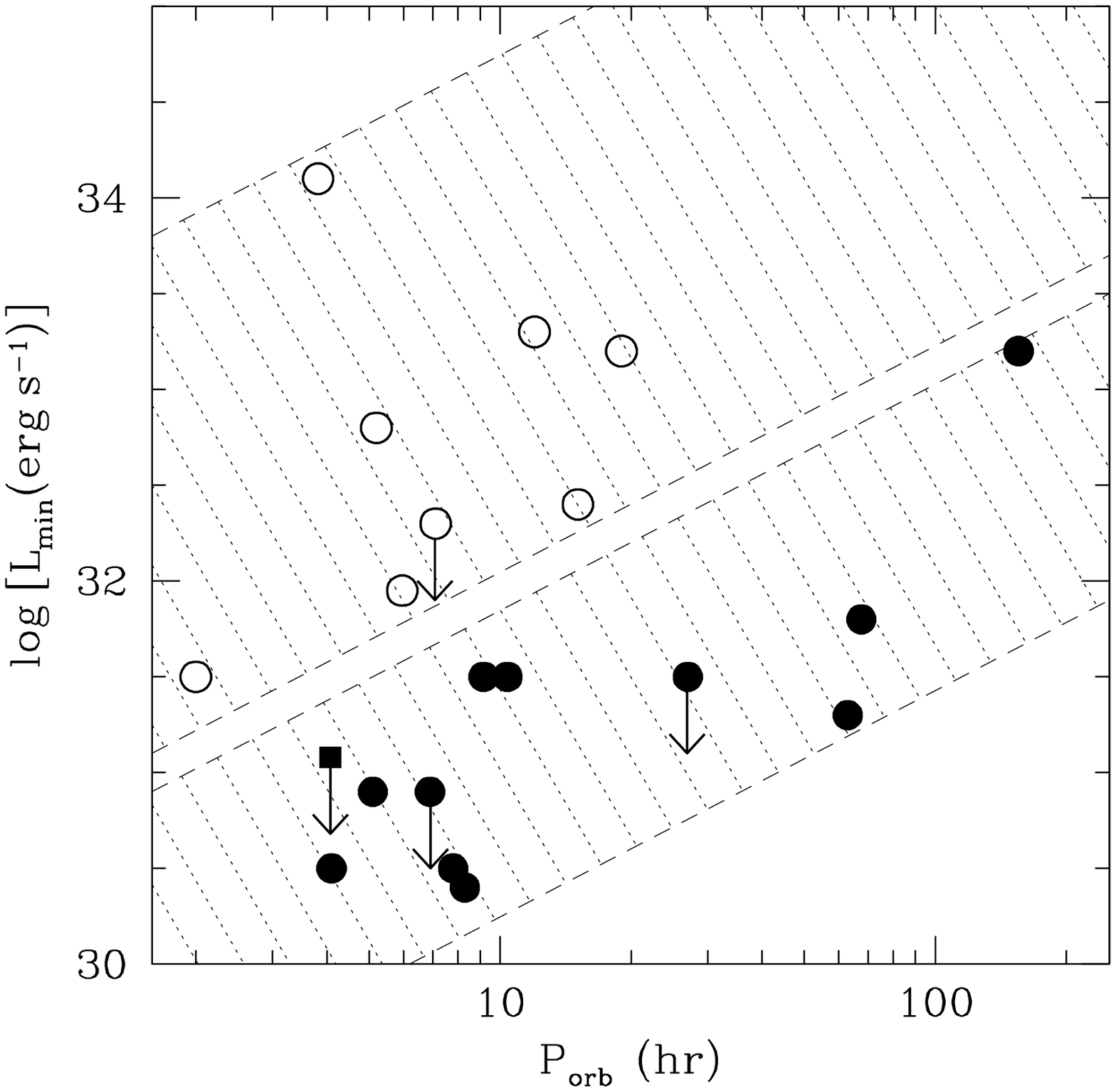] {Same as Figure \ref{fig:3} but without the
Eddington scaling.  In this representation, BH LMXBs are on average a
factor $\sim100$ fainter than NS LMXBs with similar orbital
periods.\label{fig:5}}

\figcaption[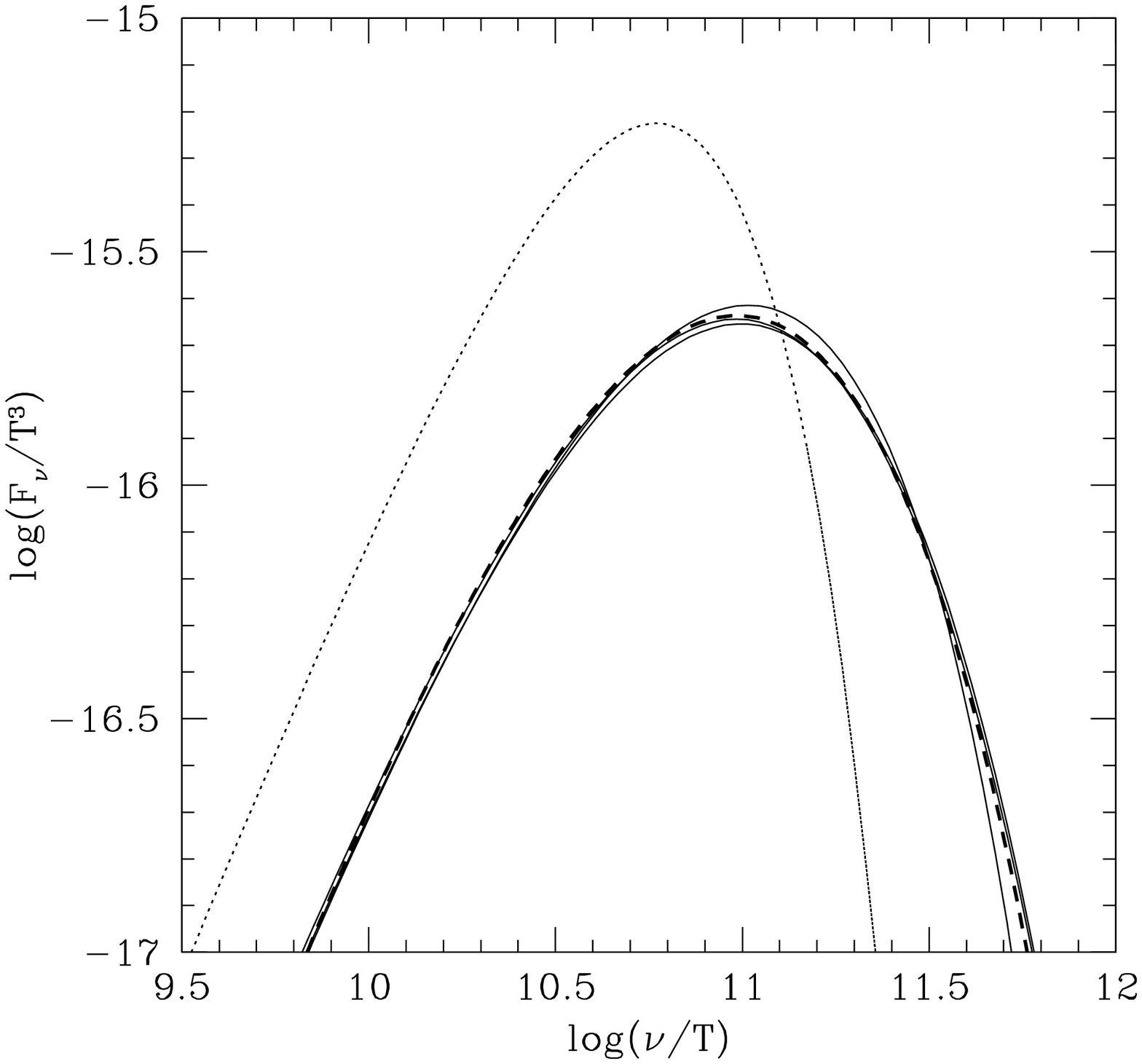] {Scaled emergent flux vs.\ scaled frequency for
various ``standard'' atmosphere models. Solid curves are for $\log g =
14.1$ and (from top down) for $\log\Teff= 6.0, 5.5, $and$ 5.0$.
Dotted curve is the blackbody flux.  Dashed curve is the approximation
of equation (\eqref{gbr17}).\label{fig:6}}

\figcaption[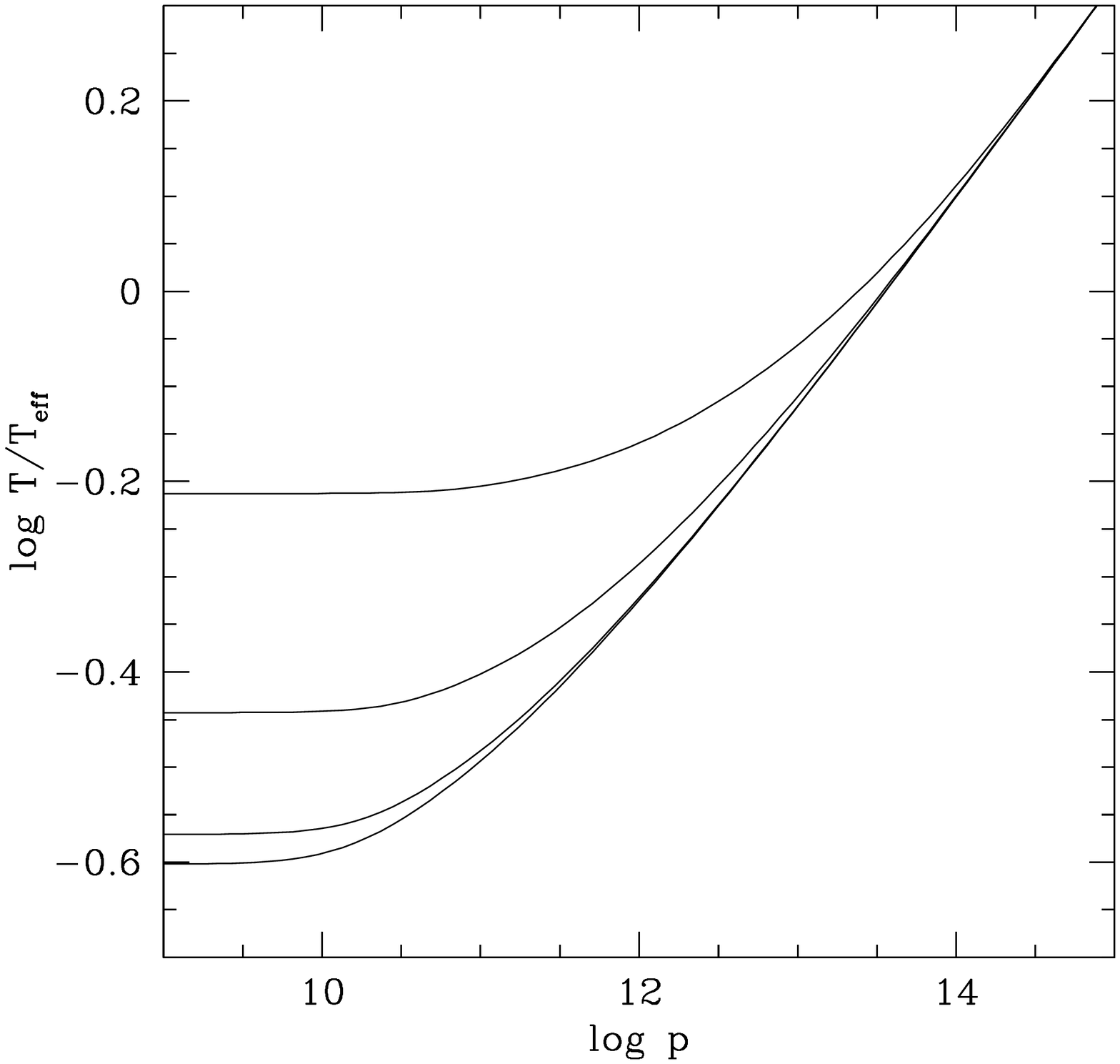] {Temperature versus pressure for models with
$\log g = 14.654$ and $\log\Teff=6.0$. The curves are, from bottom to
top, for $\muc=0$, 0.1094, 0.3522, and 0.6383.\label{fig:7}}

\figcaption[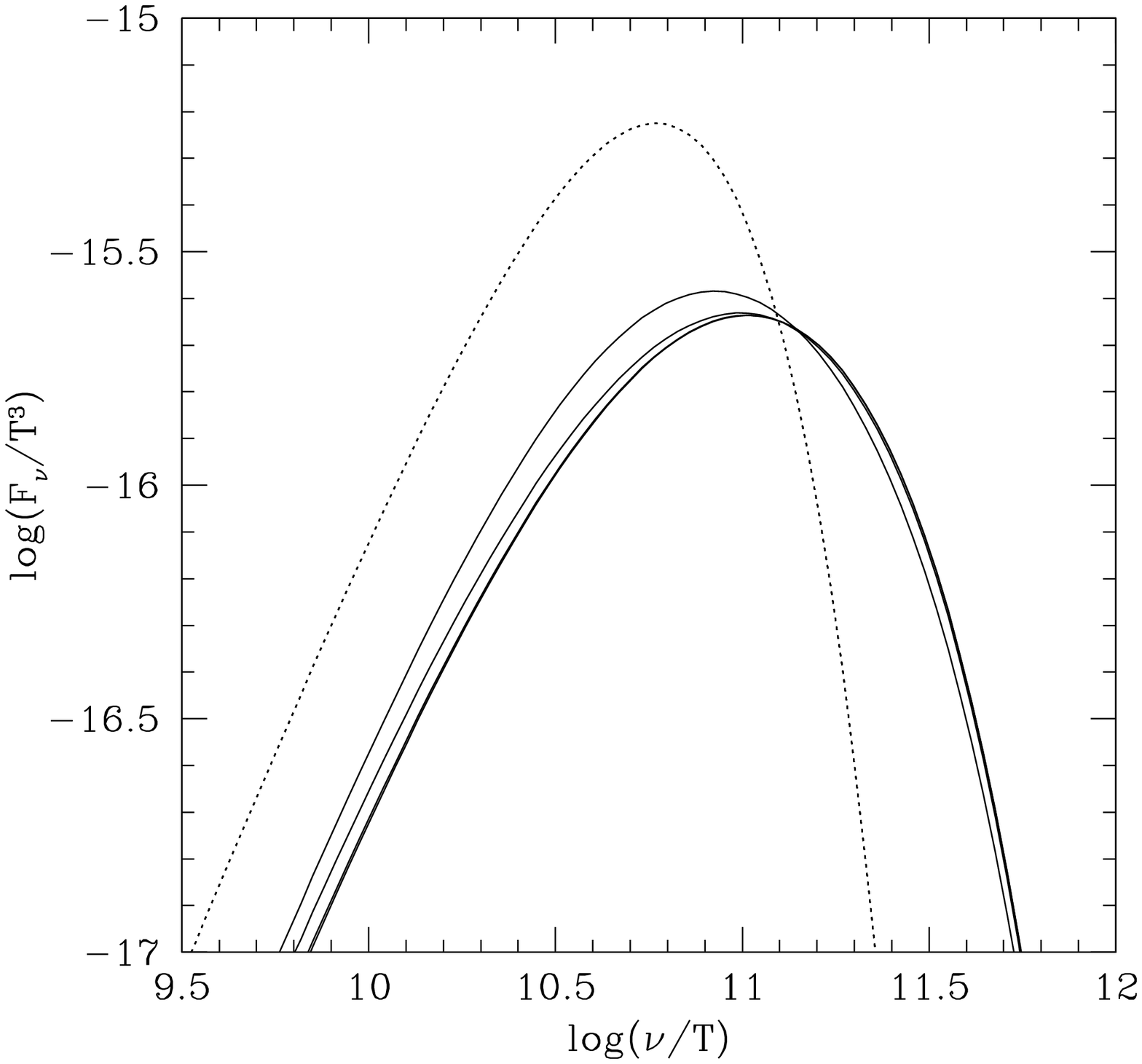] {Solid curves are spectra for the models in
Figure \ref{fig:8}.  At frequencies well below the peak, the curves
are, from bottom to top, for $\muc=$ 0, 0.1094, 0.3522, and 0.6383.
Note that the first two are barely distinguishable.  Dotted curve is
the blackbody flux.\label{fig:8}}

\newpage
\begin{figure}
\figurenum{1}
\plotone{f1.eps}
\caption{ }
\end{figure}

\newpage
\begin{figure}
\figurenum{2}
\plotone{f2.eps}
\caption{ }
\end{figure}

\newpage
\begin{figure}
\figurenum{3}
\plotone{f3.eps}
\caption{ }
\end{figure}

\newpage
\begin{figure}
\figurenum{4}
\plotone{f4.eps}
\caption{ }
\end{figure}

\newpage
\begin{figure}
\figurenum{5}
\plotone{f5.eps}
\caption{ }
\end{figure}

\newpage
\begin{figure}
\figurenum{6}
\plotone{f6.eps}
\caption{ }
\end{figure}

\newpage
\begin{figure}
\figurenum{7}
\plotone{f7.eps}
\caption{ }
\end{figure}

\newpage
\begin{figure}
\figurenum{8}
\plotone{f8.eps}
\caption{ }
\end{figure}

\end{document}